\documentstyle[preprint,tighten,pra,aps,latexsym,amssymb,floats,epsfig]{revtex}

\newcommand{\rif}[1]{(\ref{#1})}

\newcommand{\CP}[1]{{\mathbb C}{\rm P}^{#1}}

\newcommand{\eq}{\begin{equation}}
\newcommand{\feq}{\end{equation}}
\newcommand{\eqn}{\begin{eqnarray}}
\newcommand{\feqn}{\end{eqnarray}}
\newcommand{\arr}{\begin{eqnarray*}}
\newcommand{\farr}{\end{eqnarray*}}

\newcommand{\M}{{\cal M}}

\newcommand{\lp}{\left(}
\newcommand{\rp}{\right)}

\newcommand{\R}{{\mathbb R}}

\begin{document}
\tightenlines
\draft

%%%%%    GREEK ALPHABET  
\def\al{\alpha}
\def\be{\beta}
\def\ga{\gamma}
\def\de{\delta}
\def\ep{\varepsilon}
\def\ze{\zeta}
\def\io{\iota}
\def\ka{\kappa}
\def\la{\lambda}
\def\roh{\varrho}
\def\si{\sigma}
\def\om{\omega}
\def\ph{\varphi}
\def\th{\theta}
\def\te{\vartheta}
\def\up{\upsilon}
\def\Ga{\Gamma}
\def\De{\Delta}
\def\La{\Lambda}
\def\Si{\Sigma}
\def\Om{\Omega}
\def\Te{\Theta}
\def\Th{\Theta}
\def\Up{\Upsilon}

\preprint{UTF 431}

\title{M-Theory and Stringy Corrections to Anti-de~Sitter Black Holes and
Conformal Field Theories}

\author{Marco M.~Caldarelli\footnote{email: caldarel@science.unitn.it}
and Dietmar Klemm\footnote{email: klemm@science.unitn.it}\\ \vspace*{0.5cm}}

\address{Universit\`a  degli Studi di Trento,\\
Dipartimento di Fisica,\\
Via Sommarive 14\\
38050 Povo (TN)\\
Italia\\
\vspace*{0.5cm}      
and\\ Istituto Nazionale di Fisica Nucleare,\\
Gruppo Collegato di Trento,\\ Italia}

\maketitle
\begin{abstract}
We consider black holes in anti-de~Sitter space AdS$_{p+2}$ ($p = 2,3,5$),
which have hyperbolic, flat or spherical event horizons.
The ${\cal O}(\alpha'^3)$ corrections (or the leading corrections in
powers of the eleven-dimensional Planck length, in the case of M-theory
compactifications) to the black hole metrics are computed for the
various topologies and dimensions. We investigate the consequences
of the stringy or M-theory corrections for the black hole
thermodynamics. In particular, we show the emergence of a stable
branch of small spherical black holes. Surprisingly, for any
of the considered dimensions and topologies, the corrected
thermodynamical quantities turn out to coincide with those calculated
within a simplified approach, which uses only the unperturbed metric.
We obtain the corrected Hawking-Page transition temperature for black holes
with spherical horizons, and show that for $p=3$ this phase transition
disappears at a value of $\alpha'$ considerably
smaller than that estimated previously by Gao and Li. Using the AdS/CFT
correspondence, we determine the $S^1 \times S^3$
${\cal N}=4$ SYM phase diagram for sufficiently large `t Hooft coupling,
and show that the
critical point at which the Hawking-Page transition disappears
(the correspondence point of Horowitz-Polchinski), occurs at
$g_{YM}^2N \approx 20.5$. The $d=4$ and $d=7$ black hole phase diagrams
are also determined, and connection is made with the corresponding
boundary CFTs.
Finally, for flat and hyperbolic
horizons, we show that
the leading stringy or M-theory corrections do not give rise to any phase
transition. However, if the horizon is compactified to a torus $T^p$ or to a
quotient of hyperbolic space, ${\mathbb H}^p/\Gamma$,
the appearance of light winding modes around
non-contractible cycles signal new phase transitions, which in the toroidal
case have previously been discussed by Barb\'{o}n et al.~. We comment
on these phase transitions for SYM on ${\mathbb H}^p/\Gamma$ and on $T^p$,
when the moduli of the torus are taken into account.
\end{abstract}


\maketitle

\section{Introduction}

Black holes in anti-de~Sitter space have turned out to be important
tools to study large $N$ gauge theories. The connection between
black hole physics and field theory is provided by the AdS/CFT correspondence
\cite{malda,witten1,witten2,kleb1}, according to which
supergravity on AdS spaces (times some compact manifold) is dual
to certain superconformal field theories.
An interesting application of this correspondence was the observation
of Witten \cite{witten2}, who interpreted the Hawking-Page phase
transition \cite{hawkpage} from thermal five-dimensional
AdS space at low temperature to a Schwarz\-schild-AdS
black hole at high temperature, as a phase transition
in the corresponding $d=4$
super Yang-Mills theory in the limit of large `t Hooft coupling.
For the black hole, the free energy scales as $N^2$,
whereas for thermal AdS space, it is of order one. On the Yang-Mills side,
therefore, the low temperature phase is confining, whereas the high
temperature phase is deconfining, the $N^2$ scaling representing
the gluon degrees of freedom. This example feeds the hope that
the AdS/CFT correspondence finally will help us to understand
confinement in QCD \cite{ks,gubser,girard} (cf.~also \cite{kleb2}
for a review of the present status of the art).\\
Now, for AdS black holes, the presence of the negative cosmological
constant makes it possible to circumvent Hawking's theorem forbidding
non-spherical event horizon topology \cite{hawking}.
As a result, in AdS space one can have a large variety of
black holes, whose event horizons are Einstein manifolds
${\cal M}$ with positive, negative, or zero scalar curvature
\cite{amin,mann,lemos,cai,huang,vanzo,birmingham}, the latter arising
also as near-horizon limits of certain $p$-branes in
supergravity theories \cite{emparan}. If the event horizon is
diffeomorphic to $\R^p$ ($p$-brane solution),
Witten \cite{witten2} showed that there is no Hawking-Page
phase transition, i.~e.~the system is always in the high temperature
(black hole) phase. This feature was later shown by Birmingham
\cite{birmingham} to be valid also for hyperbolic black holes, i.~e.~for
those whose event horizons have negative scalar curvature.
According to the AdS/CFT dictionary,
this means that the corresponding field theory, which now lives on
$S^1 \times {\cal M}$, has no phase
transition as a function of the temperature in the large $N$ limit,
if ${\cal M}$ is hyperbolic or if ${\cal M} \approx \R^p$.
In the $\R^p$ case,
this is immediately clear from the fact that the circumference
$\beta$ of the $S^1$ factor can be scaled out by conformal invariance,
whereas for spherical black holes, leading to a CFT on $S^1 \times S^p$,
we can dispose of the circumference $\beta'$ of $S^p$ to form the
conformally invariant ratio $\beta/\beta'$ \cite{witten2}.
For hyperbolic black holes, however, the absence of any phase transition
at nonzero temperature is not so clear, as the curvature radius of
${\cal M}$ provides a characteristic scale.\\
If one wishes to go beyond the supergravity approximation, in order
to study conformal field theories on the boundary of AdS, one has
to calculate stringy corrections (or corrections in powers of
the eleven-dimensional Planck length, in the case of M-theory)
to the bulk spacetimes. This has been done in \cite{kleb3} for
the D3-, M2-, and M5-branes. (Cf.~also \cite{theisen} for the
case of D3-branes).
The motivation for doing so was that the
free energy of $d=4$
maximally supersymmetric $SU(N)$ gauge theory showed a
seeming discrepancy of a factor $3/4$
between the calculations
performed by using near-extremal D3-brane geometry, and the weak
coupling limit. (Actually, this is not a contradiction, because
the supergravity approximation is valid for {\it strong} `t Hooft coupling
$g_{YM}^2N$). The authors of \cite{kleb3} showed that the leading stringy
corrections raise this factor $3/4$. As far as the five-dimensional
Schwarz\-schild-AdS black hole is concerned, stringy corrections
to the thermodynamics
have first been computed in \cite{gaoli}, by using a simplified
perturbation theory method. (Later, the results of \cite{gaoli} have
been confirmed by a calculation of the $\alpha'$-perturbed metric
\cite{landsteiner}).
It was shown that these corrections
lower the Hawking-page transition temperature. Furthermore, the
authors of \cite{gaoli} argued that the Hawking-Page transition
eventually disappears at sufficiently small coupling. This would
then correspond to a critical point in the SYM phase diagram,
where the first order Hawking-Page transition degenerates into a
higher order phase transition.
That point was conjectured to be identified with the correspondence point of
Horowitz and Polchinski \cite{horowitz}, which (in its simplest version)
describes the transition from a highly excited string state to a black hole,
when the characteristic curvature radius at the horizon becomes of order the
string length. The entropy is in fact continuous at this correspondence
point, like in a higher order phase transition. This continuity has
been used in \cite{horowitz} to provide a statistical interpretation
of the entropy of Schwarz\-schild black holes by string states.\\
Whereas the previous papers \cite{kleb3,theisen,gaoli,landsteiner} treat the
five-dimensional case, with event horizon either diffeomorphic
to $\R^3$ or to $S^3$,
we shall extend the computations of stringy-
and M-theory corrections to arbitrary event horizon topology, and
also to the dimensions $d=4$ and $d=7$. 
In all cases, we determine
the corrected metric, and compare the thermodynamical quantities
obtained in this manner with those calculated within a perturbative
approach, which uses the uncorrected metric.
Quite interestingly, the corrected black hole entropy and free energy
will turn out to coincide with the corresponding quantities computed
with the perturbative technique. In five dimensions, this result was already
obtained in \cite{kleb3} and \cite{landsteiner} for the case of flat
or spherical event horizons. It is surprising that it holds also for
the hyperbolic case, which involves the subtraction of a nontrivial
background in computing the action. This background, unlike the AdS
vacuum used in the spherical or flat case, represents an extremal
black hole which itself receives stringy corrections.\\
Furthermore, we shall see that some interesting
new features arise, like the emergence of a stable branch of small
spherical black holes. We will determine the behaviour of the
Hawking-Page transition temperature in function of $\alpha'$ for $p=3$
(or, alternatively, in function of $g_{YM}^2N$). By using the simplified
perturbative approach mentioned above,
Gao and Li \cite{gaoli} were able to give
a rough estimate for the critical value of $\alpha'$ at which the
Hawking-Page transition disappears. By extrapolating the $T_c = T_c(\alpha')$
curve to zero temperature, they obtained $\alpha'_{crit}
\approx 0.55 l^2$, where $l$ is the AdS$_5$ radius. This value corresponds
to $g_{YM}^2N \approx 1.65$. We shall see below that a detailed analysis
yields a disappearance of the Hawking-Page transition already at
a critical value of $\alpha'$ considerably smaller, namely
$\alpha'_{crit} \approx 0.16 l^2$, corresponding to $g_{YM}^2N \approx 20.5$.
In the phase diagram, this gives rise to a critical point (the
Horowitz-Polchinski correspondence point), at which the first order
transition degenerates to higher order.
Besides, we show that there is a second critical value
$g_{YM}^2N \approx 33.4$ in the phase diagram, at which an additional
phase of small black holes appears. Taking into account the different
black hole or AdS phases, we will be able to derive the phase diagram
of SYM on $S^1 \times S^3$ for sufficiently large `t Hooft coupling.
We also determine the phase structures of the $d=4$ and $d=7$ spherical
black holes in M-theory, relevant for ${\cal N}=8$ SYM on $S^1 \times S^2$
and an exotic $(0,2)$ CFT on $S^1 \times S^5$ respectively.\\
On the other hand, in the hyperbolic
case, it will turn out that the stringy corrections do not
lead to the appearance of phase transitions, if the hyperbolic
space is simply connected. If it is compactified, instead, strings
can wrap around the non-contractible 1-cycles, and we shall briefly
comment on the effect of light winding modes on the modification
of the geometry and the possible appearance of phase transitions.
In the flat case, ${\cal M} \approx \R^p$, as already remarked above,
there can be no phase transition at any nonzero temperature. This
changes however, if we compactify the $\R^p$ down to a torus. 
(For black holes and SYM on tori cf.~\cite{martinec1,martinec2}).
The implications of these finite size effects on the emergence of
phase transitions have been studied in \cite{barbon}, and we will also
give a short comment on the more general case when the torus is not
symmetric, i.~e.~if one takes into account its moduli.\\
Finally, as further motivation to determine stringy corrections to black holes,
we would like to mention that, apart from the possibility
to study the thermodynamics of the corresponding
conformal field theories on the boundary, corrections in $\alpha'$ are
also interesting on the black hole (bulk) side,
from the point of view of a possible resolution of black hole
singularities \cite{tseytlin}.\\
The remainder of this paper is organized as follows:
To make it self-contained, we begin in section \ref{topbh} by giving
a short introduction into the geometry and thermodynamics of $d$-dimensional
topological black holes.
In \ref{comp}, some AdS compactifications
of string- and M-theory are reviewed, with emphasis
on non-trivial compactifications
involving Hopf bundles over complex projective spaces. In \ref{simpleappr},
we give the leading corrections to the supergravity actions, and
use a simple approach involving the unperturbed
metric, to calculate the leading corrections to the black hole
free energy and entropy. In section \ref{corr}, the
corrected black hole metrics
are determined in the various cases, and in \ref{thermo}
the resulting thermodynamical
quantities are discussed and compared with those obtained by
the method used in \ref{simpleappr}. Furthermore, the various phase
diagrams are determined.
In \ref{disc}, we summarize our results, and discuss them with regard to
conformal field theories.

\section{Geometry and Thermodynamics of Topological Black Holes}
\label{topbh}

One peculiarity of anti-de~Sitter space consists in the fact that the negative
cosmological constant allows for black holes with non-spherical event horizon
topology. The first works in the four-dimensional case
\cite{amin,mann,lemos,cai,huang,vanzo} have been
generalized by Birmingham \cite{birmingham} to arbitrary dimension.
The general solution in $d$ dimensions is given by the metric
\eq
ds^2 = -f(r)dt^2+f^{-1}(r)dr^2+r^2 h_{ij}(x)dx^idx^j,
\label{tbh}
\feq
where the coordinates are labeled as $x^\mu=(t,r,x^i)$,
$i=1,\dots,d-2$, and the metric $h_{ij}(x)$ describes the transverse
manifold ${\cal M}^{(d-2)}$. For (\ref{tbh}) to be a solution of
Einstein's equations with cosmological constant
\eq
\Lambda = -\frac{(d-1)(d-2)}{2l^2},
\label{cosmo}\feq
the manifold ${\cal M}^{(d-2)}$ must be Einstein,
\eq
{\cal R}_{ij} = (d-3)kh_{ij},
\feq
${\cal R}_{ij}$ denoting the Ricci tensor of $h_{ij}$, and $k$ is a
constant, which can be restricted to the values $k=0,\pm1$ without
loss of generality. $k=1,0,-1$ then corresponds to elliptic, flat,
or hyperbolic horizons respectively.
For the function $f$, one gets
\eq
f(r)=k-\frac{\eta}{r^{d-3}}+\frac{r^2}{l^2}, \label{f}
\feq
where $\eta$ is an integration constant, which is related to the
black hole mass.
Note that the manifold ${\cal M}^{(d-2)}$ is not required to
be of constant curvature, it merely has to be Einstein.
In $d=6$, for instance, we could take ${\cal M}^4 \approx S^2 \times S^2$,
which clearly can carry no metric of constant curvature, but it can
carry an Einstein metric. If, on the other hand, ${\cal M}^{(d-2)}$
is of constant curvature (which we will assume in the following,
as we want the solution with $\eta=0$ to be locally anti-de~Sitter
\cite{birmingham}
\footnote{Observe that if ${\cal M}^{(d-2)}$ is three-dimensional,
the fact that it is Einstein implies that it is of constant curvature.}),
it must be a quotient of $S^{d-2}$, $\R^{d-2}$,
or ${\mathbb H}^{d-2}$ for $k=1$, $k=0$, or $k=-1$ respectively,
where ${\mathbb H}^{d-2}$ denotes $(d-2)$-dimensional hyperbolic (Lobachevsky)
space. This means that even in the case $k=1$ we are not restricted to
the sphere, but we can also have lens spaces etc.\\
The spacetime with metric (\ref{tbh}) has an event horizon at $r=r_+$,
where $r_+$ is the largest root of $f(r)=0$. It is interesting
that for $k=-1$, one
can also have black holes for negative values of the parameter
$\eta$; the equation $f(r)=0$ has positive solutions for
\eq
\eta \ge \eta_0 = -\frac{2}{d-1}\left(\frac{d-3}{d-1}\right)^{\frac{d-3}{2}}
                  l^{d-3}. \label{eta0}
\feq
For $\eta = \eta_0$, the black hole becomes extremal, with an event horizon
at
\eq
r_+ = r_0 = \left(\frac{d-3}{d-1}\right)^{\frac{1}{2}}l. \label{r0}
\feq
For $k=-1$ and $\eta=0$, one has black holes which are locally AdS;
they represent higher-dimensional generalizations of the BTZ black hole
\footnote{Note that in the case $k=-1,\eta=0$,
one has a black hole interpretation only
if the space ${\mathbb H}^{d-2}$ has been compactified; the spacetime
with topology $\R^2 \times {\mathbb H}^{d-2}$ represents AdS seen
by a uniformly accelerated observer, with $r=r_+=l$ denoting its
acceleration horizon \cite{vanzo}.}.
The thermodynamics of the black holes presented above,
and their implications on the boundary conformal field theories
have been examined in \cite{birmingham}.
Requiring the absence of conical singularities in the euclidean metric
yields the Hawking temperature
\eq
T=\frac{(d-1)r_+^2+(d-3)kl^2}{4\pi l^2r_+}. \label{T}
\feq
The mass is given by
\eq
M=\frac{r_+^{(d-3)}}{\omega_d}\lp k+\frac{r_+^2}{l^2}\rp +
  \frac{\eta_0}{\omega_d}\delta_{k,-1}, \label{M}
\feq
where
\eq
\omega_d = \frac{16\pi G_d}{(d-2)V_{d-2}}.
\feq
Here $V_{d-2}$ denotes the volume of ${\cal M}^{(d-2)}$, which we
assumed to be compact; otherwise one would have to define the mass per
unit volume. The term proportional to the Kronecker delta in (\ref{M})
corresponds to the correct choice of background, such that the
Arnowitt-Deser-Misner (ADM) mass is a positive, concave function of the
black hole's temperature \cite{vanzo}. This background is the one with
$\eta=0$ (i.~e.~locally AdS)
for $k=0,1$, but for $k=-1$, one has to choose the extremal
black hole with $\eta=\eta_0$ as reference background \cite{vanzo}.
The euclidean action reads
\eq
I_d^{(0)} = \frac{V_{d-2}\beta}{16\pi G_d}\left[-\frac{r_+^{d-1}}{l^2} +
          kr_+^{d-3} + 2\frac{d-2}{d-1}\left(\frac{d-3}{d-1}
          \right)^{\frac{d-3}{2}}l^{d-3}\delta_{k,-1}\right], \label{act0bh}
\feq
yielding the entropy
\eq
S=\frac{V_{d-2}}{4G_d}r_+^{d-2}. \label{entr}
\feq
In (\ref{act0bh}) and (\ref{entr}), the horizon radial coordinate $r_+$
can be expressed as a function of the temperature via
\eq
r_+(T) = \frac{2\pi l^2 T}{d-1}\left[1 \pm \left(1 - \frac{(d-1)(d-3)k}
         {(2\pi T l)^2}\right)^{\frac{1}{2}}\right]. \label{r+}
\feq
Here the lower sign makes sense only in the spherical ($k=1$) case,
where two different values of $r_+$ are possible for a given temperature,
the branch of small black holes being unstable \cite{hawkpage}.
The free energy of the black hole as well as that of the
boundary conformal field theory is then given by $F = I_d^{(0)}/\beta$,
and is easily seen to scale as $V_{d-2}T^{d-1}$ in the limit of
high temperatures, independently of the curvature $k$.
(In the planar case, $k=0$, this scaling law is valid for all
temperatures).

\section{Topological Black Holes in String- and M-Theory} \label{comp}

In this section we shall review how the black holes described above arise as
solutions of the low energy effective actions of M-theory or string theory
for special values of $d$.
We shall first see how the topological black holes, for $d=4$,
emerge in low-energy limit of M-theory and type IIA string theory as exact
solutions of $D=11$, $N=1$ supergravity and type IIA supergravity respectively.
Emparan \cite{emparan} considered $d=4$ topological black holes in their
M-theory context. The solution of the corresponding low-energy effective
action, i.~e.~of $D=11$ supergravity, reads
\eq
ds^2=-f(r)dt^2+f^{-1}(r)dr^2+r^2d\sigma_k^2+4l^2d\Omega_7^2\label{M4}
\feq
where $f(r)=k-\eta/r+r^2/l^2$, $d\sigma^2_k$ is the metric on $S^2$,
the two-torus or ${\mathbb H}^2$ for $k=1,0,-1$ respectively, and
$d\Omega^2_7$ denotes the metric on $S^7$. The four-form field strength
is $F_{\mu\nu\rho\sigma}=(3/l)\,\epsilon_{\mu\nu\rho\sigma}$, and
corresponds to the Freund-Rubin ansatz for the eleven dimensional manifold.
The associated 3-form potential $A_{[3]}$ can be chosen to be
$A_{t\theta\phi}=-3r/l$, the other components being zero.
These objects also describe the near-horizon limit of
non-extremal M2-branes wrapped around compact surfaces of arbitrary
genus.\\
Proceeding with a Kaluza-Klein compactification on a circle of M-theory, one
obtains type IIA superstring theory \cite{witten3}.
In particular, compactification of
\rif{M4} leads to a topological black hole in type IIA supergravity.
One observes first that the odd-dimensional sphere $S^{2n-1}$
can be written as a $U(1)$ bundle over $\CP{n-1}$; it is then possible to
perform the Kaluza-Klein reduction on the $U(1)$ direction of the Hopf
fibres \cite{dlp98}. Hence, we write the $S^7$ factor of the metric 
\rif{M4} as
\eq
d\Omega_7^2=d\Sigma_6^2+(dz+\bar{\cal A}_Mdz^M)^2,
\feq
where $d\Sigma_6^2$ is the standard Fubini-Study metric on 
$\CP3$, and the Kaluza-Klein vector potential $\bar{\cal A}$ 
has field strength $\bar{\cal F}$ given by $\bar{\cal F}=2J$, where $J$
is the K\"ahler form on $\CP3$. The coordinate on the fiber
is $z$ and has period $4\pi$, while the coordinates $z^M$, $M=1\dots6$,
parametrize $\CP3$.
We can now compactify M-theory along the $S^1$ direction
parametrized by the coordinate $z$. The resulting field configuration will
solve the equations of motion of type IIA supergravity; the
corresponding metric is described by the line element
\eq
ds_{IIA}^2=2^{1/4}\lp-f(r)dt^2+\frac{dr^2}{f(r)}+r^2d\sigma_k^2+4l^2d
\Sigma_6^2\rp,
\feq
and represents a direct product of a four-dimensional black hole and
$\CP3$. The dilaton is constant, and assumes the value
$\phi=\frac32\ln2$.
The Ramond/Ramond (RR) one-form $A_{[1]}$ can be read directly from the 
mixed components $g_{\mu z}$; it is non-vanishing only on
$\CP3$, $A_{[1]M}=l\bar{\cal A}_M$.
The remaining fields of IIA supergravity, the Neveu-Schwarz/Neveu-Schwarz
(NSNS) two-form $B_{[2]}$ and the RR three-form $C_{[3]}$, are determined
by dimensional reduction of the three-form $A_{[3]}$ of $D=11$ supergravity,
\eq
A_{[3]}=C_{[3]}+B_{[2]}\wedge l dz,
\feq
from which we deduce that the antisymmetric tensor field $B_{[2]}$ vanishes
while the RR three-form coincides with the three-potential of $D=11$
supergravity, $C_{[3]t\theta\phi}=-\frac{3r}l$. Note that the
compactification of IIA supergravity on $\CP3$ breaks some supersymmetry;
the reason is that some Killing spinors are described by
D0-branes, and the full supersymmetry is restored only at type IIA superstring
theory level\cite{dlp97}.\\
Modifying slightly the Freund-Rubin ansatz we obtain seven-dimensional
topological black holes in D=11 supergravity. We ask again the eleven
dimensional manifold to be a direct product ${\cal M}_7\times{\cal M}_4$,
but now we take $\M_7$ lorentzian and $\M_4$ euclidean. We require the field
$F$ to be non-vanishing only on $\M_4$, where it takes the value
$F_{mnpq}=(6/l)\,\epsilon_{mnpq}$.
This yields an Einstein space $\M_4$ of positive curvature, and we take it to
be the four-sphere, which is maximally symmetric. Other choices are
possible, at the cost of breaking some of the supersymmetries. The $\M_7$
factor is an Einstein space of negative curvature, for example any of the
seven-dimensional topological black holes \rif{tbh}. As a result we
have 
\eq
ds^2=-f(r)dt^2+f^{-1}(r)dr^2+r^2 h_{ij}^{(5)}(x)dx^idx^j+\frac14\,l^2
d\Omega^2_4,
\feq
where $f(r)=k-\eta/r^4+r^2/l^2$,
and $h^{(5)}_{ij}$ is the metric of a five-dimensional compact Einstein
space of positive, zero or negative curvature according to the sign of $k$.
Observe that we cannot obtain a seven dimensional black hole in IIA
supergravity from this solution, because $S^4$ has even dimension and
hence cannot be written as a $U(1)$ bundle.\\
Let us turn now to five dimensional topological black holes. In type IIB
supergravity there is a self-dual five-form $F$ that can be used for a
Freund-Rubin like ansatz \cite{birmingham2}. Taking 
\eq
F_{\mu_1\dots\mu_5}=\frac4l\epsilon_{\mu_1\dots\mu_5}\,,\qquad
F_{m_1\dots m_5}=-\frac4l\epsilon_{m_1\dots m_5},
\feq
and all the other fields zero, the ten-dimensional spacetime 
can become the direct product of a
five-dimensional topological black hole and a five-dimensional positive
curvature Einstein space, which represents a
solution of the IIB supergravity equations of
motion. The metric reads
\eq
ds^2=-f(r)dt^2+f^{-1}(r)dr^2+r^2 h_{ij}^{(3)}(x)dx^idx^j+l^2\,d\Omega^2_5,
\feq
where $f(r)=k-\eta/r^2+r^2/l^2$, and $h^{(3)}_{ij}$ denotes the metric of a
three-dimensional compact Einstein space of
positive, zero or negative curvature, according to the sign of $k$.\\
Furthermore, if the compact space is $S^5$, we can untwist
it \cite{dlp98}, and obtain a five-dimensional black hole times a compact space
$\CP2\times U(1)$ as solution of type IIA supergravity.
Let us see in detail how this works. The sphere $S^5$ is a $U(1)$
bundle over $\CP2$ and its metric is
\eq
ds^2(S^5)=l^2d\Sigma^2_4+l^2(dz+\bar{\cal A})^2,
\feq
where $d\Sigma_4^2$ is the metric on the ``unit'' $\CP2$, and 
$d\bar{\cal A}=2J$, $J$ being the K\"ahler form on $\CP2$.
Now, performing a T-duality along $z$, we obtain a solution of type IIA
supergravity \cite{bho95}. The dual metric reads
\eq
ds^2_{IIA}=-f(r)dt^2+f^{-1}(r)dr^2+r^2 h_{ij}^{(3)}(x)dx^idx^j+l^2d\Sigma_4^2
           +l^2 dz^2,
\feq
and is of the form $TBH_5\times\CP2\times S^1$. The RR 4-form of IIB theory
is mapped by T-duality on the RR 3-form of IIA theory,
\eq
dC_{[3]}=4l^3\,\Sigma_4,
\feq
where $\Sigma_4$ denotes the volume-form on $\CP2$,
and the Kaluza-Klein vector component of the ten-dimensional IIB metric
is mapped on the 2-form $B_{[2]}$ of the NSNS sector of IIA supergravity,
\eq
dB_{[2]}=2l\,J\wedge dz,
\feq
while the RR 1-form and the dilaton $\phi$ vanish. Observe that $\CP2$ does
not admit any spin structure. As a consequence the spectrum of Kaluza-Klein
excitations in the $\CP2\times S^1$ compactification of type IIA
supergravity contains no fermions.\\
Finally, the oxidation of the $TBH_5\times\CP2\times S^1$ solution of
IIA supergravity yields a five-dimensional topological black hole solution
of D=11 supergravity
\begin{eqnarray}
ds^2_{11}&=&ds^2_{TBH_5}+l^2d\Sigma_4^2+dz_1^2+dz_2^2\,,\\
F&=&4l^2\,\Sigma_4-2l J\wedge dz_1\wedge dz_2\,,
\end{eqnarray}
which is of the form $TBH_5\times\CP2\times T^2$.\\
In fact, more general five-dimensional AdS black hole solutions can be
found, of the form 
$TBH_5\times{\cal M}^6$, where ${\cal M}^6$ is Einstein-K\"ahler and F is
the product of the K\"ahler form with itself \cite{dn}. 
In the case of M-theory compactifications to $d=4,7$,
one can choose less symmetric compactification
spaces, substituting for example the spheres with lens spaces, or
taking other Einstein manifolds. The resulting solutions have less
supersymmetries, and could show relevant to study CFT with less than
maximal supersymmetry \cite{halyo}.\\
In conclusion, we have obtained several topological black hole
solutions. In D=11 supergravity we have $TBH_4\times S^7$,
$TBH_7\times S^4$ and $TBH_5\times\CP2\times T^2$ black holes, in
type IIA supergravity we obtained $TBH_4\times\CP3$ and
$TBH_5\times\CP2\times S^1$ black holes, while in IIB supergravity we
have the $TBH_5\times S^5$ black holes.

\section{Stringy and M-theory Corrections to Free Energy and Entropy 
from the Unperturbed Metric} \label{simpleappr}

\subsection{The Modified Supergravity Actions} \label{action}

Our aim is to determine effects of massive string states on the
black hole geometry. In the low energy effective action, these
massive string states manifest themselves as
higher derivative curvature
terms. In type IIB supergravity, the lowest correction is
of order $\alpha'^3R^4$, where $R$ denotes the Riemann tensor.
Taking this into account,
the  string-corrected IIB supergravity action reads
in Einstein frame
\cite{zanon,gross}\footnote{We impose self-duality of the five-form
field strength $F_5$ after the equations of motion are derived.}
\eq
I_{10} = -\frac{1}{16\pi G_{10}}\int d^{10}x\sqrt{g}\left[R - \frac{1}{2}
         (\partial\phi)^2 - \frac{1}{4\cdot 5!}(F_5)^2 + \ldots + \gamma
         e^{-\frac{3}{2}\phi}W + \ldots\right], \label{action10}
\feq
where
\eq
\ga = \frac{1}{8}\zeta(3)(\alpha')^3,
\feq
and the dots stand for terms depending on further antisymmetric
tensor field strengths and dilaton derivatives. Due to a field
redefinition ambiguity \cite{gross,tseytlin86}, there exists a scheme
where $W$ depends only on the Weyl tensor,
\eq
W = C^{hmnk}C_{pmnq}C_h^{\,\,rsp}C^q_{\,\,rsk} + \frac{1}{2}
    C^{hkmn}C_{pqmn}C_h^{\,\,rsp}C^q_{\,\,rsk}. \label{W}
\feq
For eleven-dimensional supergravity, the situation is somewhat
different, as there is no parameter $\alpha'$ in M-theory.
However, there are corrections in powers of the
Planck length $\ell_{11}$, leading to the
effective action (cf.~\cite{kleb3} and references therein)
\eq
I_{11} = -\frac1{2\kappa_{11}^2}\int d^{11}x\,\sqrt{-g}
\left[R+2\kappa_{11}^{4/3}\xi W+\ldots\right], \label{action11}
\feq
where $\xi=2\pi^2/3$, and the eleven-dimensional Newton constant reads in
terms of the Planck length $\ell_{11}$
\eq
8\pi G_{11} = \kappa_{11}^2=2^4\pi^5\ell_{11}^9.
\feq
As will be argued shortly, for the problem at hand one can use the
four- or seven-dimensional analogue of (\ref{W}) for the term $W$ occuring
in (\ref{action11}).
For further purpose, it will
be useful to define also in eleven dimensions
an expansion parameter $\gamma$, which is now given by
$\gamma=2\kappa^{4/3}_{11}\xi$, so that the $W$-dependent term
in the corrected actions has the prefactor $\gamma$ in both ten and
eleven dimensions.

\subsection{Free Energy and Entropy}

In order to get a first clue of the string- or M-theory corrections
to thermodynamical quantities like free energy or black hole entropy,
one can plug the {\it unperturbed} metric into the expression (\ref{W}),
and then calculate the corrected action. This approach corresponds essentially
to first order perturbation theory. It was used in \cite{kleb3} to
calculate $\alpha'$ corrections to the throat approximation
of D3-branes, and Planck length
corrections to that of
M2- and M5-branes, in order
to determine the free energy behaviour of conformal field theories
beyond the supergravity approximation in the AdS/CFT correspondence.
The authors of \cite{kleb3} showed that in the D3-brane case, this
simple procedure yields exactly the same result as a calculation
involving the perturbed metric. For the
five-dimensional $k=1$ Schwarz\-schild-AdS
black hole, Gao and Li \cite{gaoli} used the same approach to show
that the Hawking-Page phase transition temperature is lowered
by the first $\alpha'$ correction. Later, Landsteiner \cite{landsteiner}
computed explicitely the perturbed black hole metric, and obtained
the same black hole entropy as that supplied by the perturbative approach of
\cite{gaoli}. In section \ref{thermo} we will see that this coincidence
extends also to the hyperbolic case ($k=-1$), and to the dimensions
$d=4,7$. This result is rather puzzling, and it would be
interesting to get a better physical understanding of it.\\
In the following, we will calculate the corrected action for the
$d$-dimensional topological black hole spacetimes (\ref{tbh}),
where we shall concentrate on the cases $d = p+2 = 4,5,7$, corresponding to
the well-known AdS compactifications of M-theory
or of type IIB string theory. In calculating the term (\ref{W})
quartic in the Weyl tensor, one actually should take the full
ten- or eleven-dimensional metric, i.~e.~the product of (\ref{tbh})
and the metric on a suitably chosen sphere. In the case of D3-branes
in the throat approximation, the full ten-dimensional expression
for $W$ coincides with the one calculated using only the five-dimensional
part without five-sphere \cite{kleb3}. This fact is related to a
vanishing scalar curvature of the full ten-dimensional metric \cite{kleb3}.
An argument which justifies the use of the $(p+2)$-dimensional analogue
of (\ref{W}) also in the case of M-branes down the throat, was given in
\cite{kleb3}. The same argument applies also if the branes are wrapped
on Einstein spaces ${\cal M}^p$, i.~e.~in our case. Therefore,
in the following, we will use the $d=p+2$ dimensional analogue
of (\ref{W}), and compute it for the metric (\ref{tbh}).
$W$ is then given by
\eq
W = d_{p+2}\left[\frac{\eta(d-2)(d-1)}{2r^{d-1}}\right]^4, \label{WBH}
\feq
where we defined the coefficients
\eq
d_4 = \frac{1}{18}, \qquad d_5 = \frac{5}{36}, \qquad d_7 = \frac{292}{1125}.
\feq
Upon dimensional reduction on a suitable sphere, the correction to
the actions (\ref{action10}) resp.~(\ref{action11}) becomes
\eq
\delta I_d = -\frac{1}{16\pi G_d}\int d^dx \sqrt{g_d}\gamma W, \qquad
             d=4,5,7, \label{deltaI}
\feq
where
\eqn
\frac{1}{16\pi G_5} &=& \frac{{\mathrm{Vol}}(S^5)}{16\pi G_{10}} =
                        \frac{\pi^3 l^5}{16\pi G_{10}}, \nonumber \\
\frac{1}{16\pi G_4} &=& \frac{{\mathrm{Vol}}(S^7)}{16\pi G_{11}} =
                        \frac{\pi^4 (2l)^7/3}{16\pi G_{11}}, \\
\frac{1}{16\pi G_7} &=& \frac{{\mathrm{Vol}}(S^4)}{16\pi G_{11}} =
                        \frac{8\pi^2 (l/2)^4/3}{16\pi G_{11}}. \nonumber
\feqn
(Note that in our conventions the radii of the four- and seven spheres
are given by $l/2$ and $2l$ respectively). In (\ref{deltaI}), $\gamma$
is defined by $\gamma = \zeta(3)(\alpha')^3/8$ for $d=5$, and by
$\gamma = 4\pi^2\kappa_{11}^{4/3}/3$ for $d=4,7$.
Plugging (\ref{WBH}) into (\ref{deltaI}),
one can compute the stringy or M-theory corrections. 
In doing so,
one has to take care of an important subtlety. In the calculation of
the unperturbed action, one performs a subtraction of a suitably
chosen background in order to cancel divergences. Now, as already
observed in \ref{topbh}, for the cases
$k = 0$ or $k=1$ this background is (at least locally) anti-de Sitter
space, which receives no corrections, as it has $\eta=0$. For the
hyperbolic black hole ($k=-1$) however, the correct background is the
extremal black hole with $\eta=\eta_0$ (cf.~(\ref{eta0})).
In calculating the corrected action, we must therefore subtract
the background value, which gives a nonvanishing contribution
in the hyperbolic case, as can be seen from (\ref{WBH}).
In this way, we get
\eq
\delta I_d = -\frac{\gamma V_{d-2}\beta}{16\pi G_d}d_{p+2}\frac{(d-2)^4
             (d-1)^3}{3}\left[\frac{r_+^{d-9}}{16l^8}\left(kl^2+
             r_+^2\right)^4 - \delta_{k,-1}\frac{l^{d-9}}{(d-1)^4}
             \left(\frac{d-3}{d-1}\right)^{\frac{d-9}{2}}\right],
             \label{deltaIcorr}
\feq
where $r_+$ is a function of the temperature according to (\ref{r+}).
The correction to the free energy is then $\delta F_d = \delta I_d/\beta$,
which again scales as $V_{d-2}T^{d-1}$ for high temperatures, independently
of the curvature $k$. One immediately checks (after adapting the conventions),
that for $k=0$, (\ref{deltaIcorr}) reproduces the results of \cite{kleb3},
and for $d=5,k=1$ it reduces to the corrected action obtained in \cite{gaoli}.
We further observe that for $k=0,1$ the correction to the action is
always negative.
The entropy correction can be computed by $S = -\frac{\partial F}{\partial T}$,
which yields
\eqn
\delta S_d &=& \frac{\gamma}{G_d}d_{p+2}\frac{(d-2)^4\pi^3}{3(d-3)}\left(
             \frac{2\pi l^2}{d-1}\right)^{d-5}V_pT^p[1 + (1-x)^{\frac{1}{2}}]^
             {d-5}\cdot \nonumber \\
           & & \cdot \left[\frac{d-2}{d-3} - \frac{1}{d-3}(1-x)^{\frac{1}
               {2}}\right]^3[3 + (d-6)(1-x)^{-\frac{1}{2}}]\label{deltaS},
\feqn
where we defined
\eq
x = \frac{(d-1)(d-3)k}{(2\pi T l)^2}.
\label{x}\feq
Note that the $k=-1,\eta=0$ locally AdS black holes, which have Hawking
temperature $T=1/(2\pi l)$, do not receive any stringy or M-theory
corrections to their entropy up to order $\alpha'^3$.
(At least not in this simplified
calculation, but we shall see later that a computation involving the
perturbed metric gives the same result; besides, for $\eta=0$ the metric does
not receive any corrections).
This suggests that maybe they represent exact solutions
of string- or M-theory, similarly to the BTZ black hole, which has
been shown to be an exact solution of string theory by Horowitz and Welch
\cite{welch}.

\section{Corrected Black Hole Metrics}
\label{corr}

In this section we shall compute the corrections to the metric of the
$TBH_4\times S^7$ and $TBH_7\times S^4$ black holes in M-theory
and of the $TBH_5\times S^5$ solution in type IIB superstring theory.
We shall proceed in a unified way, leaving the dimension $d$ of the black
hole arbitrary, and carrying out the perturbative expansion
in $\gamma$, defined in section \ref{action}.\\
To compute the perturbed solution, following \cite{kleb3}, we write the
reduced action for a metric of the form
\eq
ds^2=H(r)^2\lp K^2(r)\,dt^2+P^2(r)\,dr^2+l^2\,d\sigma_k^2\rp,
\feq
where $d\sigma_k^2$ is the unit metric of the transverse elliptic, flat or
hyperbolic manifold for $k=1$,$0$ and $-1$ respectively.
Using the symmetry of the ansatz, we can reduce the corrected actions
\rif{action10} and \rif{action11} to one dimension in a unified way, and we
obtain the effective action
\footnote{For the subtleties concerning the self-duality of $F_5$ in the
IIB supergravity case, we refer to \cite{kleb3}. Furthermore, a more
general ansatz should actually be used, allowing also the radius
of the compactification
space to vary \cite{theisen}. However, in \cite{kleb3} an argument is
given that shows that such a generalization
does not yield additional corrections to the free
energy; this has been explicitly verified in some cases in
\cite{theisen,landsteiner}. Hence we shall use the simplified ansatz with
fixed compactification space radius in the following.}
\eq
I=-\frac{\beta V_{d-2}}{16\pi G_d}\int_{r_+}^{\infty}
\!\!\sqrt{g}\lp R-2\Lambda+\gamma W\rp dr,
\feq
where the expansion parameter $\gamma$ is given in section \ref{action} for
IIB superstring theory and M-theory, $\beta$ is the period in euclidean time,
$V_{d-2}$ is the volume of the (unit) transverse manifold and $\Lambda$ is
the cosmological constant defined in \rif{cosmo}. In the IIB supergravity
case, the dilaton $\phi$ appears in the full action \rif{action10}. However
on the unperturbed solution it vanishes, and the leading order
perturbations of the dilaton do not mix with the metric perturbations.
Hence, in computing the corrections to the metric, we shall simply forget
about the dilaton; we will add some comments on its leading order
perturbation at the end of this section.\\
The invariant $W$ can be computed from the Weyl tensor and reads
\eq
W=\frac{d_{p+2}}{H^8}\left[\frac1{KP}\lp\frac{K'}P\rp'-\frac k{l^2}\right],
\feq
where the primes denote derivatives with respect to $r$.
To simplify the action, it is convenient to set \cite{kleb3}
\eq
H(r)=\frac{r}l\,,\qquad K(r)=e^{a(r)+4b(r)}\,,\qquad P(r)=e^{b(r)}.
\feq
Substituting the ansatz in the action we obtain
\begin{eqnarray}
&\sqrt{g}\lp R-2\Lambda\rp&=-2r^{d-3}e^{a+3b}
\left[r(a''+4b'')+(d-1)(a'+3b')+ra'^2+12rb'^2\phantom{\frac2d}\right.
\nonumber\\
&&\left.+7ra'b'+\frac{(d-4)(d-1)}{2r}\right]
+(d-2)\frac{r^{d-2}}{l^2}\left[(d-3)k+(d-1)\frac{r^2}{l^2}\right]e^{a+5b}
\end{eqnarray}
for the zero order term, and
\eq
\sqrt{g}\,W=d_{p+2}\frac{l^6}{r^{8-d}}e^{a-3b}
\lp a''+4b''+a'^2+7a'b'+12b'^2-\frac k{l^2}e^{2b}\rp^4
\feq
for the first corrections to the lagrangian.
The Euler-Lagrange equations corresponding to the reduced action yield
the equations of motion for $a(r)$ and $b(r)$,
\begin{eqnarray}
21-5d-2ra'+\left[\frac{5r^2}{l^2}\lp k(d-3)+(d-1)\frac{r^2}{l^2}\rp
+\frac\gamma{l^6}w_b(r)\right]e^{2b}&=&0,\nonumber\\
5-d+2rb'+\left[\frac{r^2}{l^2}\lp k(d-3)+(d-1)\frac{r^2}{l^2}\rp
+\frac\gamma{l^6}w_a(r)\right]e^{2b}&=&0,\label{eqmotion}
\end{eqnarray}
where we have defined
\begin{eqnarray}
w_a(r)&=&-\frac{9\eta^3}{4l^4r^8}\lp56r^3+64kl^2r-67l^2\eta\rp\,,\\
w_b(r)&=&-\frac{9\eta^3}{4l^4r^8}\lp224r^3+264kl^2r-279l^2\eta\rp\,.
\end{eqnarray}
We look now for first-order corrections to the metric. To this end, we first
expand the functions $a(r)$ and $b(r)$ in $\gamma$,
\eq
a(r)=a_0(r)+\gamma a_1(r)+\ldots\,,\qquad b(r)=b_0(r)+\gamma b_1(r)+\ldots\,,
\feq
where $a_0$ and $b_0$ are determined by the unperturbed solution,
\eq
a_0(r)=\frac12\ln\left[\frac{r^6}{l^6}f^5(r)\right],\qquad
b_0(r)=-\frac12\ln\left[\frac{r^2}{l^2}f(r)\right],
\feq
$f(r)$ being the lapse function defined in \rif{f}.
Obviously, $a_0$ and $b_0$, which describe the unperturbed AdS black
holes, solve the equations of motion \rif{eqmotion} at zero order in $\gamma$.
Expanding the equations to first order in $\gamma$ and solving them, one
finds the ${\cal O}(\gamma)$ corrections to the metric,
\begin{eqnarray}
a_1(r)&=&-\frac{(d-1)^3(d-2)^2}{96l^2r^{3d-5}f(r)}d_{p+2}\eta^3
\left[-104(2d-1)-16(13d-14)\frac{kl^2}{r^2}+9(17d-6)\frac{l^2\eta}
{r^{d-1}}\right.\nonumber\\
&&\left.+5\lp24(2d-1)+48(d-1)\frac{kl^2}{r_+^2}-(37d-14)\frac{l^2\eta}
{r_+^{d-1}}\rp\lp\frac{r}{r_+}\rp^{2d-2}\right],
\end{eqnarray}
and
\begin{eqnarray}
b_1(r)&=&\frac{(d-1)^3(d-2)^2}{96\,l^2r^{3d-5}f(r)}d_{p+2}\eta^3
\left[
-24(2d-1)-48(d-1)\frac{kl^2}{r^2}+(37d-14)\frac{l^2\eta}{r^{d-1}}
\right.\nonumber\\
&&\left.
+\lp24(2d-1)+48(d-1)\frac{kl^2}{r_+^2}-(37d-14)\frac{l^2\eta}{r_+^{d-1}}
\rp\lp\frac{r}{r_+}\rp^{2d-2}\right].
\end{eqnarray}
The integration constants in $a_1(r)$ and $b_1(r)$ have been chosen such that
the corrections vanish asymptotically, where the corrected metric must
match the locally AdS metric. Furthermore, the position of the
horizon is not modified by the corrections, which are finite for
$r\rightarrow r_+$.\\
It is interesting to note that the $\gamma$-corrections vanish for
$\eta=0$. This becomes plausible recalling that in this case the spacetimes
are (at least) locally AdS. Using supersymmetry arguments, it was argued in
\cite{kr98} that ${\mathrm{AdS}}_4\times S^7$, ${\mathrm{AdS}}_7\times S^4$
or ${\mathrm{AdS}}_5\times S^5$ are exact vacua of M-theory or IIB
string theory respectively. In the case $\eta=0, k=-1$ these arguments
do not apply anymore, as spacetime represents now a
black hole with Hawking temperature $T=1/(2\pi l)$, if the transverse
space ${\mathbb H}^p$ has been compactified. Clearly a black hole
with non-vanishing temperature cannot be supersymmetric \cite{ck98}.
Presumably, the $\eta=0, k=-1$ black holes are nevertheless
exact solutions of string theory in much the same way as the BTZ black
hole \cite{welch}.\\
Finally, in the IIB supergravity case, one should also consider the
corrections to the dilaton field. Reducing the complete action
\rif{action10} to $d=5$ dimensions, one obtains the equations of motion for
the leading order corrections $\phi_1(r)$ to the dilaton,
$\left(\sqrt{g}\,\phi_1'\right){}'=3\gamma\sqrt{g}\,W$,
and the corrections to the dilaton can be chosen to be 
\eq
\phi_1(r)=\gamma\int_{+\infty}^r\frac{45\eta^4}{2r_+^{12}}
\left[\frac{r^{12}-r^{12}_+}{r^{15}\,f(r)}\right]\,dr.
\feq
This integral can be performed analytically, but the result is quite
complicated and we shall not need it in the following.
In fact, the result can be found in
\cite{kleb3} in the $k=0$ case and in \cite{landsteiner} for $k=1$;
in the hyperbolic case the result is quite similar. The relevant
point is that the solution is regular everywhere, and the dilaton, as well
as the string coupling $g_s$, decreases
towards the horizon.

\section{Thermodynamics of the perturbed solution} \label{thermo}

Having now the corrections to the metric, we shall analyse their effects on the
thermodynamics, and compare the corrected free energy with the one obtained
from the unperturbed solution in section \ref{simpleappr}.
The temperature of the perturbed black hole is obtained
from the absence of conical singularities in the euclidean section,
\begin{eqnarray}
T(r_+)&=&\frac{(d-1)r_+^2+(d-3)kl^2}{4\pi l^2r_+}-\gamma d_{p+2}
\frac{(d-2)^2(d-1)^3}{192\pi r_+^7}\lp k+\frac{r_+^2}{l^2}\rp^3\nonumber\\
&&\times\left[(d^2+d-18)k-(d-1)(14-d)\frac{r_+^2}{l^2}\right]
+{\cal O}(\gamma^2).
\label{temp}
\end{eqnarray}
The first term of (\ref{temp}) is the temperature of the unperturbed
black hole \rif{T}, and the second term comes from the order $\gamma$ 
correction to the metric. For the toroidal black hole we have $k=0$ and the
temperature reads
\eq
T_{k=0}(r_+)=\frac{(d-1)r_+}{4\pi l^2}\lp1+\frac{\gamma d_{p+2}}{48l^6}
(d-1)^3(d-2)^2(14-d)\rp+{\cal O}(\gamma^2).
\label{Ttorus}
\feq
The behaviour of the temperature as a function of the position of the horizon
remains linear, but the coefficient is increased by the 
$\gamma$-corrections. In the spherical and hyperbolic cases, the
temperature approaches asymptotically \rif{Ttorus}.\\
One can invert \rif{temp} and obtain the black hole radius $r_+$ as
a function of the temperature $T$. We use the unperturbed radius
$r_{+(0)}$ defined in \rif{r+} as
\eq
r_{+(0)}(T) = \frac{2\pi l^2 T}{d-1}\left[1 + \sqrt{1-x}\right],
\feq
where $x$ is given by \rif{x}.
The radius of the horizon is then\footnote{In the $k=1$ case,
$r_+(T)$ describes only the branch of larger black holes.}
\begin{eqnarray}
r_+(T)=r_{+(0)}(T)&&\left[1+\gamma
\frac{d_{p+2}}{48l^6}(d-2)^2(d-1)^3\,
\frac{(r_{+(0)}^2(T)+kl^2)^3}{r_{+(0)}^6(T)}\right.\nonumber\\
&&\left.\times\frac{(d-1)(d-14)r_{+(0)}^2(T)+(d^2+d-18)kl^2}
{(d-1)r_{+(0)}^2(T)-(d-3)kl^2}\right].
\label{rt}\end{eqnarray}
We now wish to compute the corrected euclidean action of the black hole, as
well as the corrected free energy. As usual in AdS space, the surface terms of
the action do not contribute, but there is an infrared divergence coming
from the bulk contribution. This is regularized setting a cut-off in the
radial integration
and subtracting the action of an appropriate background. As mentioned
in section \ref{topbh}, in the spherical and
toroidal cases the appropriate background
is the metric with $\eta=0$. We have shown that this background receives no
correction. The hyperbolic case, however, is more
subtle, as its background \rif{eta0} gets corrections in $\gamma$.
In fact, as can be readily verified from Eq.
\rif{temp}, its formal temperature becomes negative. The right background
to subtract is the one whose {\em corrected} temperature \rif{temp}
vanishes. The position of the horizon of this corrected black hole is
obtained setting $T=0$ in Eq. \rif{rt},
\eq
r_+(T=0)=l\sqrt{\frac{d-3}{d-1}}\left[
1+\gamma\frac{d_{p+2}}{2 l^6}\frac{(d-2)^2(d-1)^3}{(d-3)^4}(3d-10)\right].
\feq
The regularized action then reads
\eq
I_{\rm reg}=\beta I(R)-\beta' I_0(R),
\label{ireg}\feq
where $I_0(R)$ is the action evaluated on the appropriate background with
the radial cut-off at $R$. The periodicity in the euclidean time of the
background $\beta'$ must match that of the black hole, in such a way that
the proper length of the circle which it parametrizes at radius $R$ is the
same as for the black hole \cite{witten2}. This yields
\eq
\beta'=\beta\left[1+\frac{(\eta_0-\eta)l^2}{2R^{(d-1)}}+{\cal O}\lp
\frac1{R^{(d+1)}}\rp\right].
\feq
At this point, we can safely eliminate the cut-off from \rif{ireg} by
sending $R$ to infinity, obtaining the black hole action
\begin{eqnarray}
I=-\frac{V_{d-2}\beta}{16\pi G_d}&&\left[
\frac{r_+^{d-1}}{l^2}-kr^{d-3}_+
+\gamma\frac{d_{p+2}(d-1)^3(d-2)^2}{48l^8r_+^{9-d}}(r_+^2+kl^2)^3
\left((11d-10)r^2_+-(5d-22)kl^2\right)\right.
\nonumber\\
&&\left.
-\lp2\frac{d-2}{d-1}\lp\frac{d-3}{d-1}\rp^{\frac{d-3}2}l^{d-3}
+\gamma\frac{d_{p+2}(d-2)^4}{3(d-1)}\lp\frac{d-3}{d-1}\rp^{\frac{d-9}2}
l^{d-9}\rp\delta_{k,-1}\right].\label{azcorr}
\end{eqnarray}
The free energy is then given by $F=I/\beta$.
Inserting Eq.~\rif{rt} into the action
\rif{azcorr} and dividing by the inverse temperature $\beta$, we finally
obtain the free energy $F(T)$ of the black hole in function of the
temperature,
\begin{eqnarray}
F(T)=-\frac{V_{d-2}}{16\pi G_d}
&&\left[\frac{r_{+(0)}^{d-1}(T)}{l^2}-kr_{+(0)}^{d-3}(T)+\frac\gamma{l^6}
\frac{d_{p+2}(d-2)^4(d-1)^3}{48l^2r_{+(0)}^{9-d}(T)}
\left(r_{+(0)}^2(T)+kl^2\right)^4\right.\nonumber\\
&&\left.
-\lp2\frac{d-2}{d-1}\lp\frac{d-3}{d-1}\rp^{\frac{d-3}2}l^{d-3}
+\gamma\frac{d_{p+2}(d-2)^4}{3(d-1)}\lp\frac{d-3}{d-1}\rp^{\frac{d-9}2}
l^{d-9}\rp\delta_{k,-1}
\right].
\end{eqnarray}
The corrected action \rif{azcorr}, written in function of $r_+$,
does not coincide with the expression \rif{deltaIcorr} + \rif{act0bh},
obtained in a naive way. However, the free energy as a function of the
temperature computed from the perturbed metric
coincides with that obtained from \rif{deltaIcorr}, as was first observed
in the cases $d=5$, $k=0$ in \cite{kleb3} and $d=5$, $k=1$ in
\cite{landsteiner}.
As a consequence, also the correction to the entropy obtained from the
perturbed metric coincides with the correction \rif{deltaS} computed from
the unperturbed metric.
This fact is particularly surprising in the $k=-1$ case, where the
background used in the regularization receives itself modifications, and
the subtraction has been done with respect to a different background.

\begin{figure}[htb]
\centerline{\epsfig{file=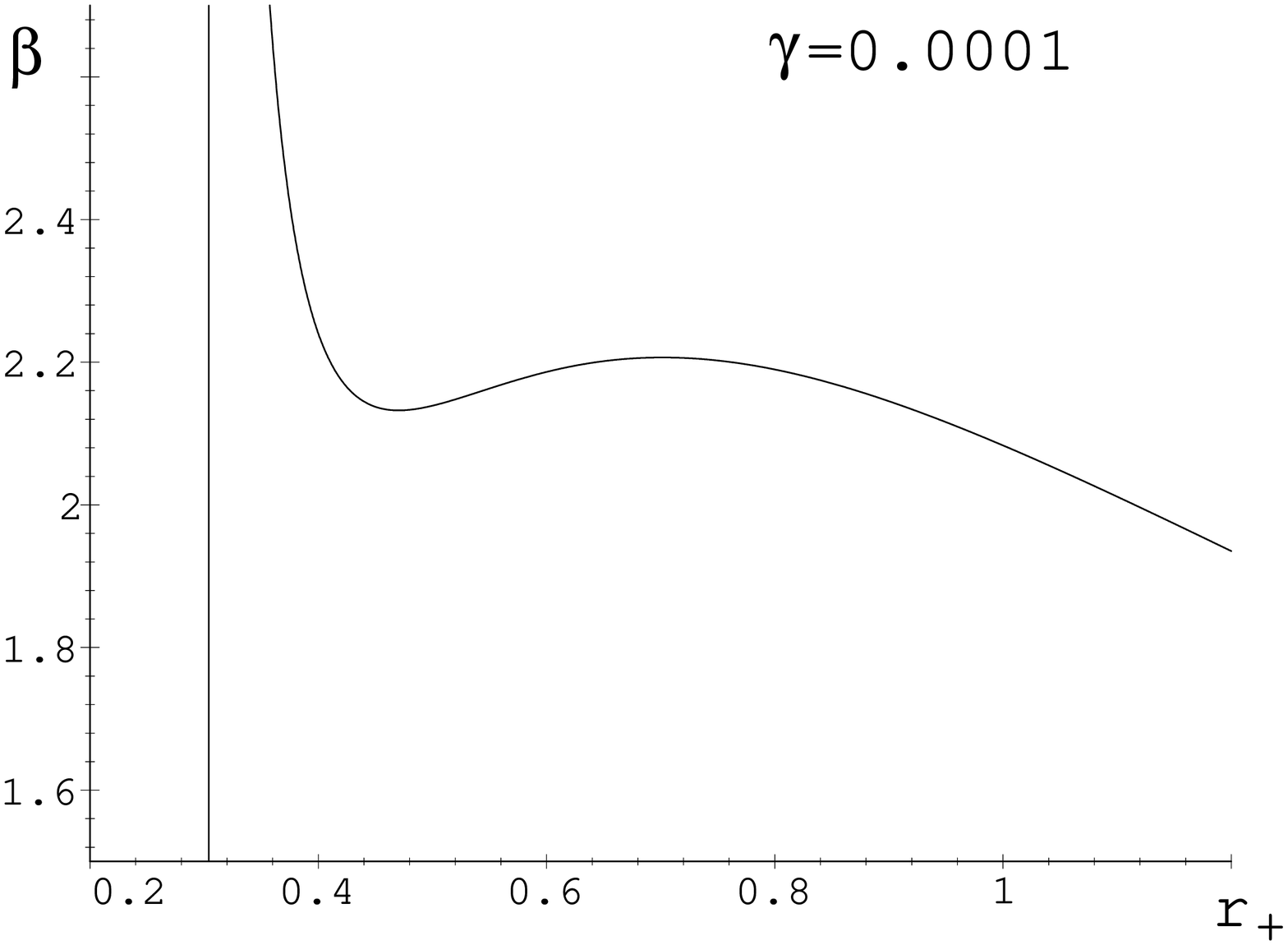,angle=-90,width=0.5\linewidth}
\epsfig{file=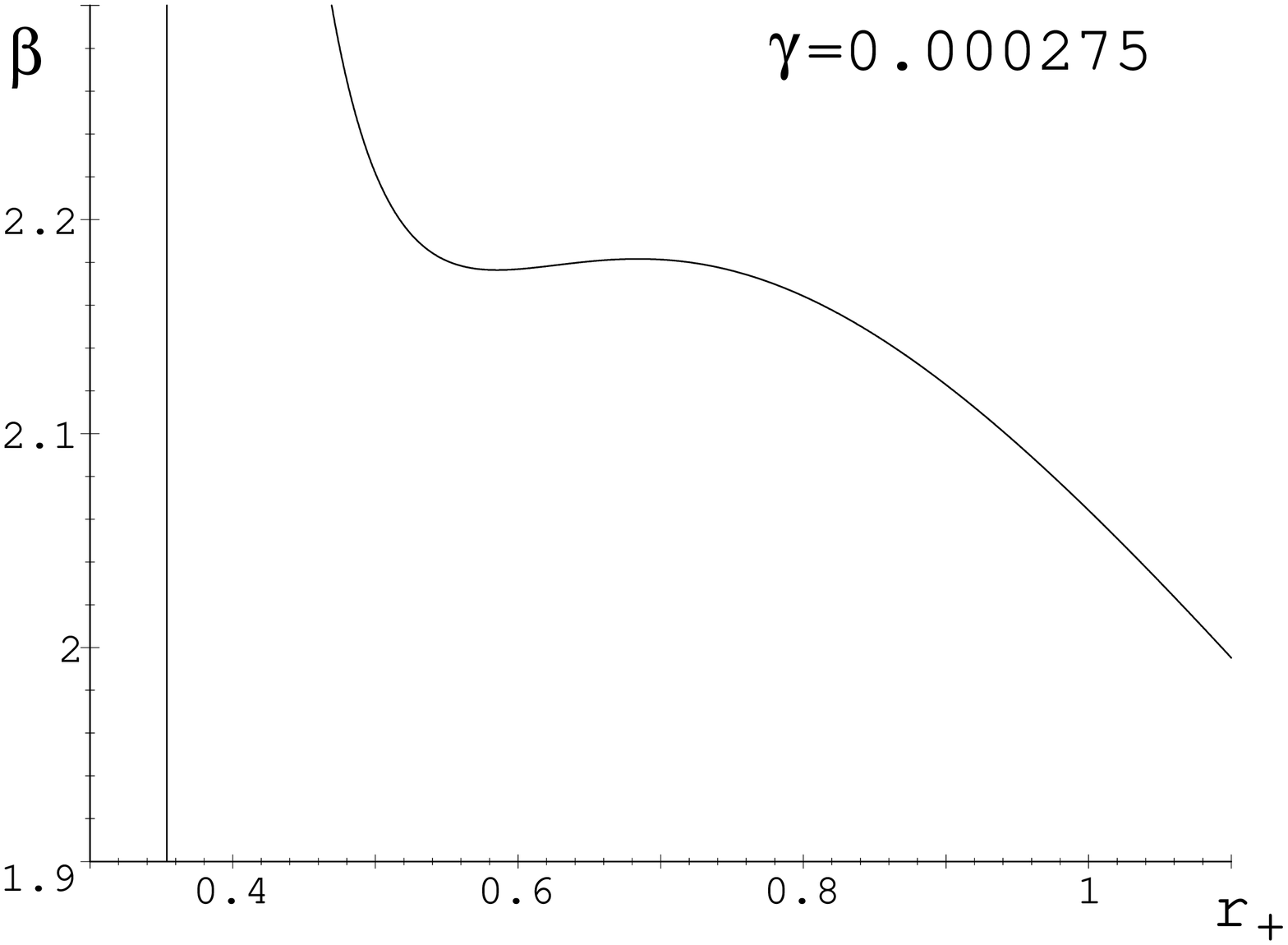,angle=-90,width=0.5\linewidth}}
\centerline{\epsfig{file=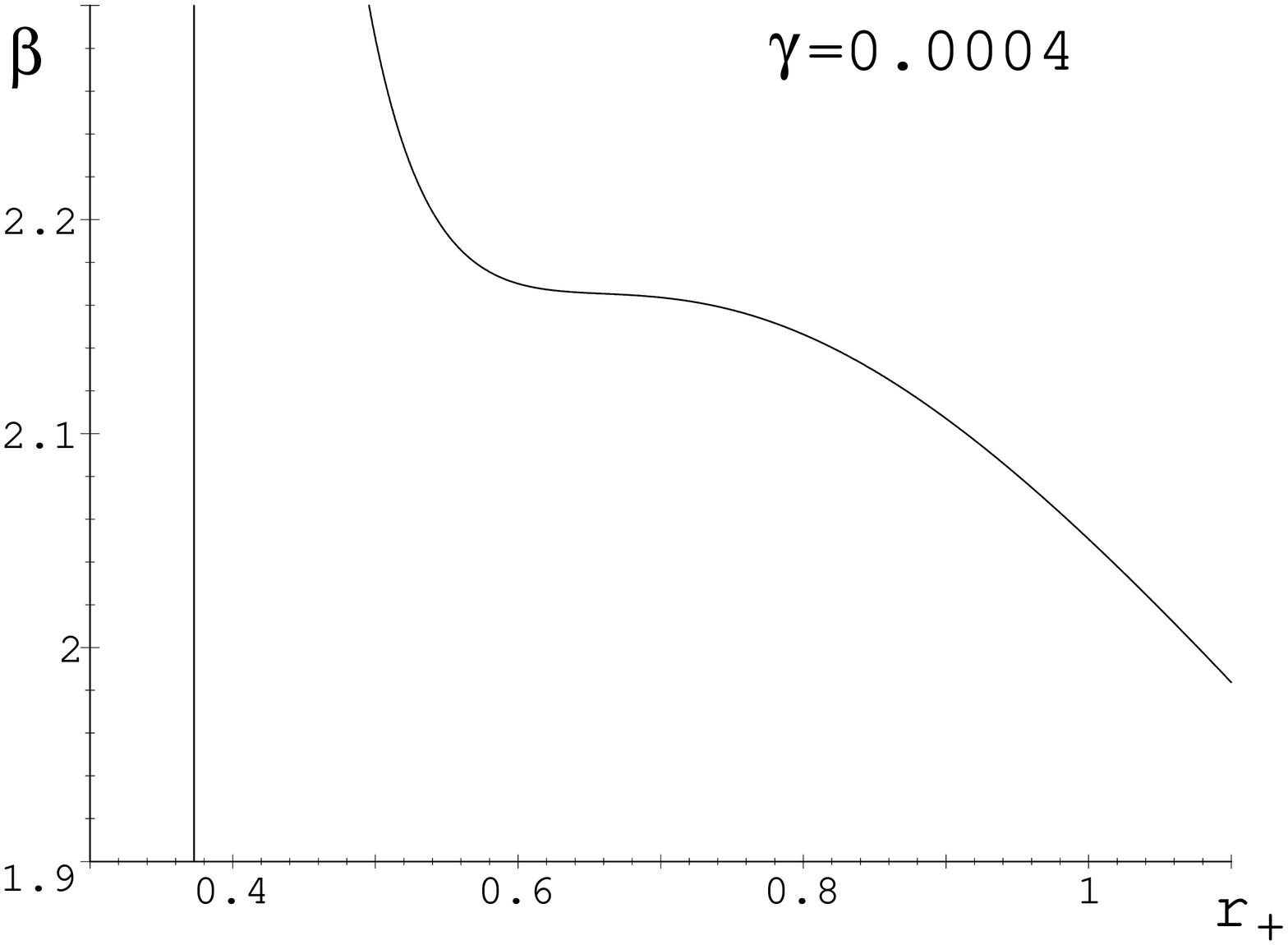,angle=-90,width=0.5\linewidth}
\epsfig{file=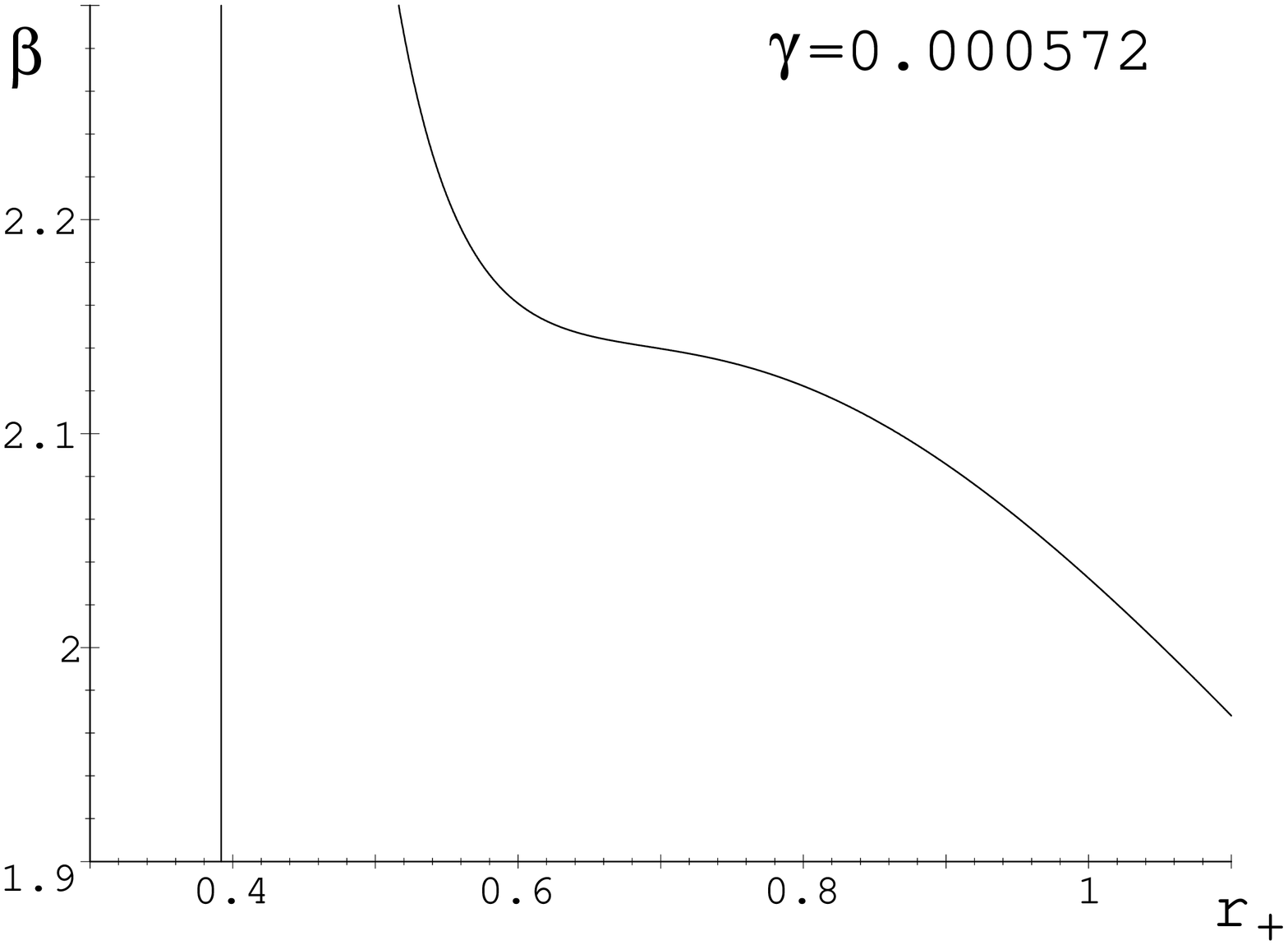,angle=-90,width=0.5\linewidth}}
\caption{{\sf Inverse temperature $\beta$ in function of the horizon
position $r_+$ for spherical black holes. A new branch of small stable
black holes appears. For small $\gamma$ there are three branches; then, at
$\gamma_{infl}\approx 0.0004$ a point of inflection appears,
and for larger $\gamma$
the small and large black hole branches merge. (Here and in the following
figures $\gamma$ is given in units of $l^6$ and $r_+$ and $\beta$ are given
in units of $l$.)}}
\label{fig:br}
\end{figure}

\begin{figure}[htb]
\centerline{\epsfig{file=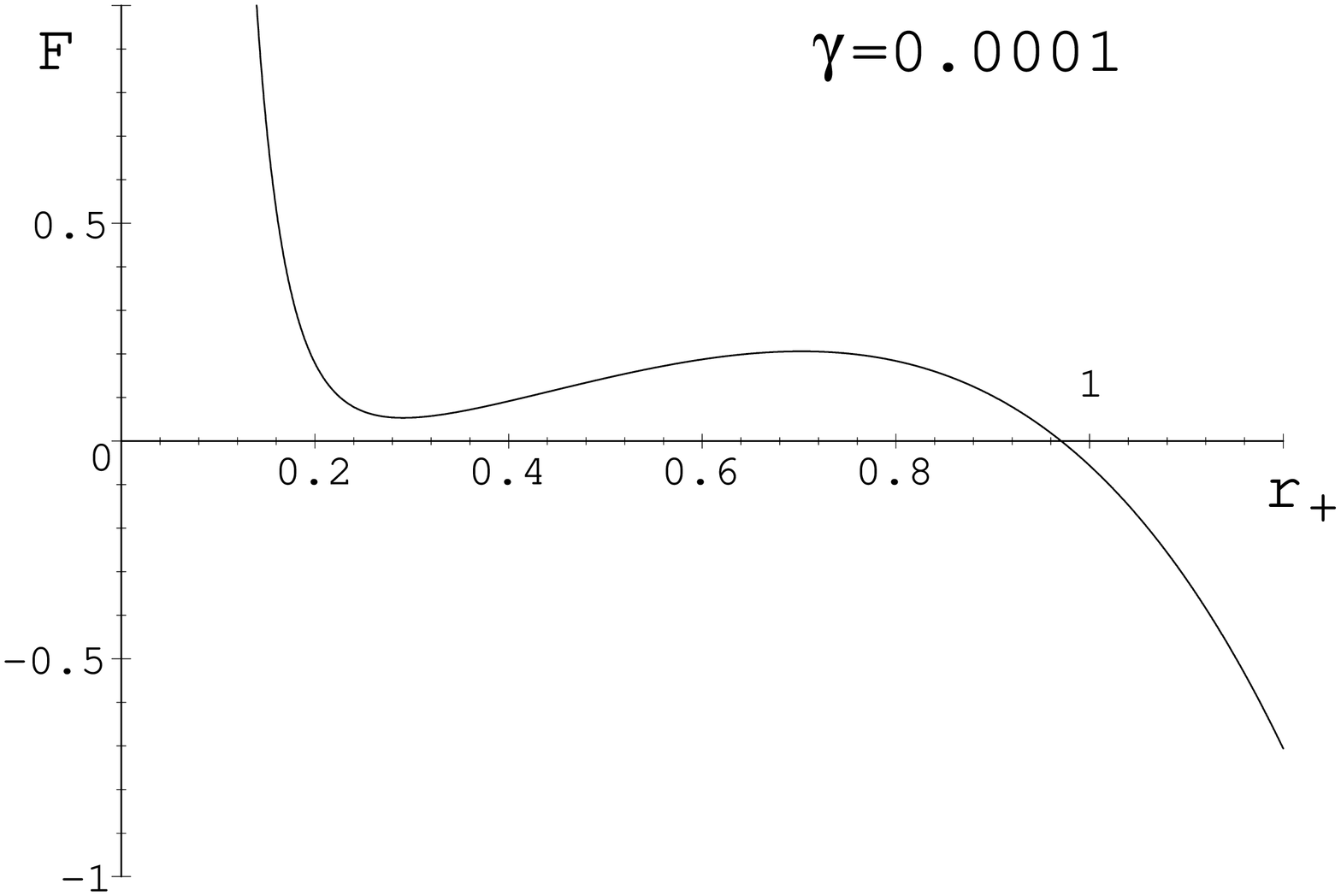,angle=-90,width=0.5\linewidth}
\epsfig{file=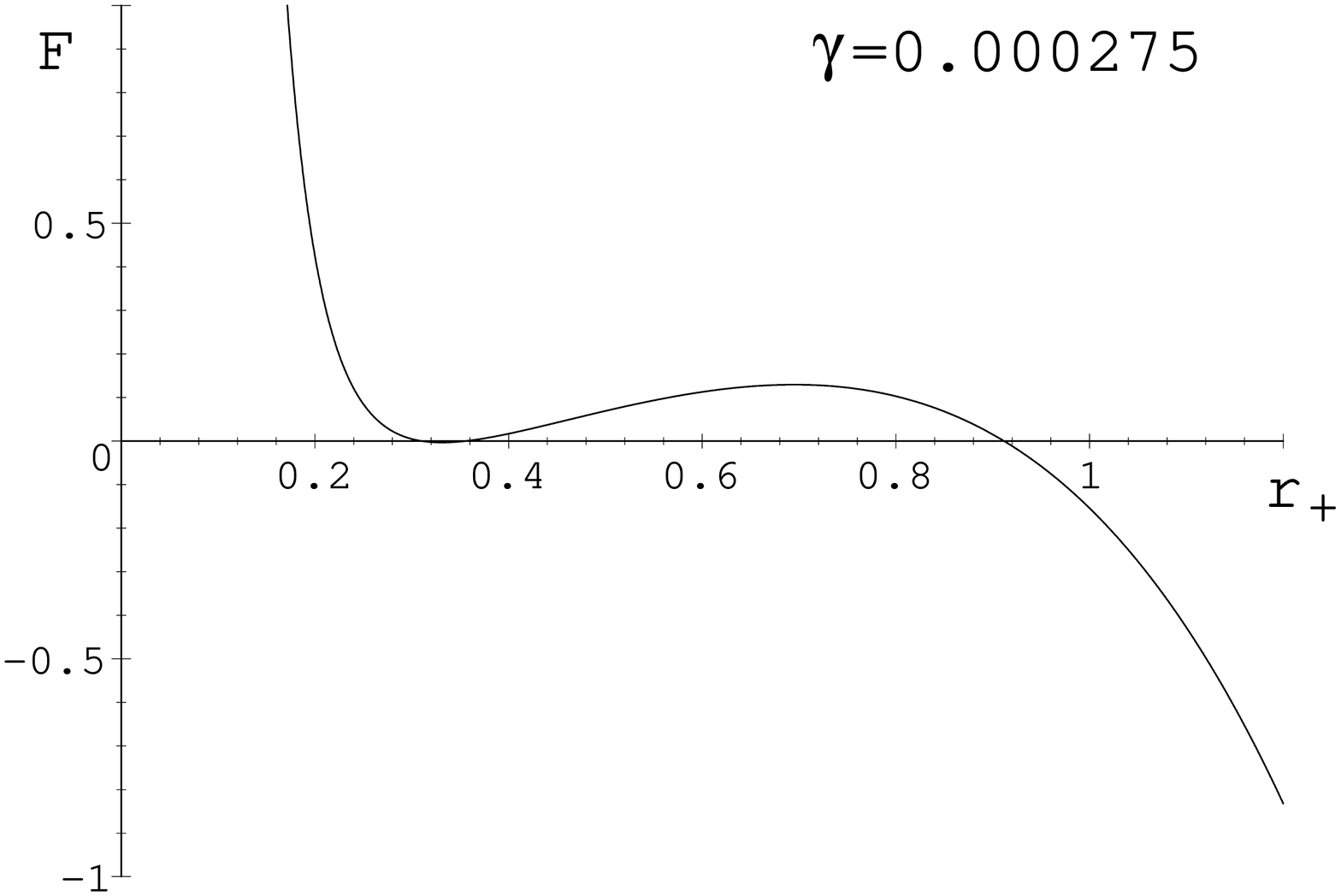,angle=-90,width=0.5\linewidth}}
\centerline{\epsfig{file=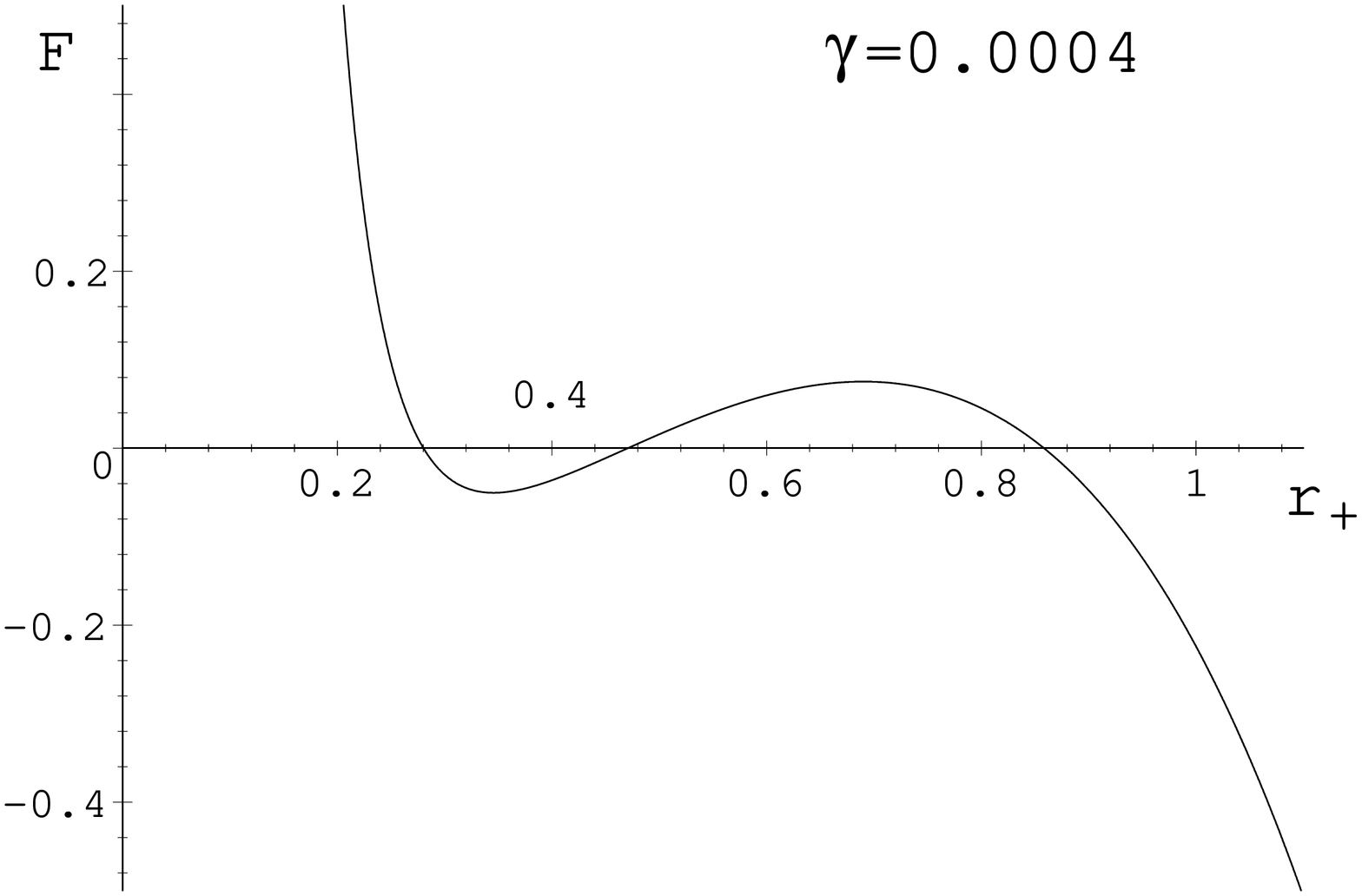,angle=-90,width=0.5\linewidth}
\epsfig{file=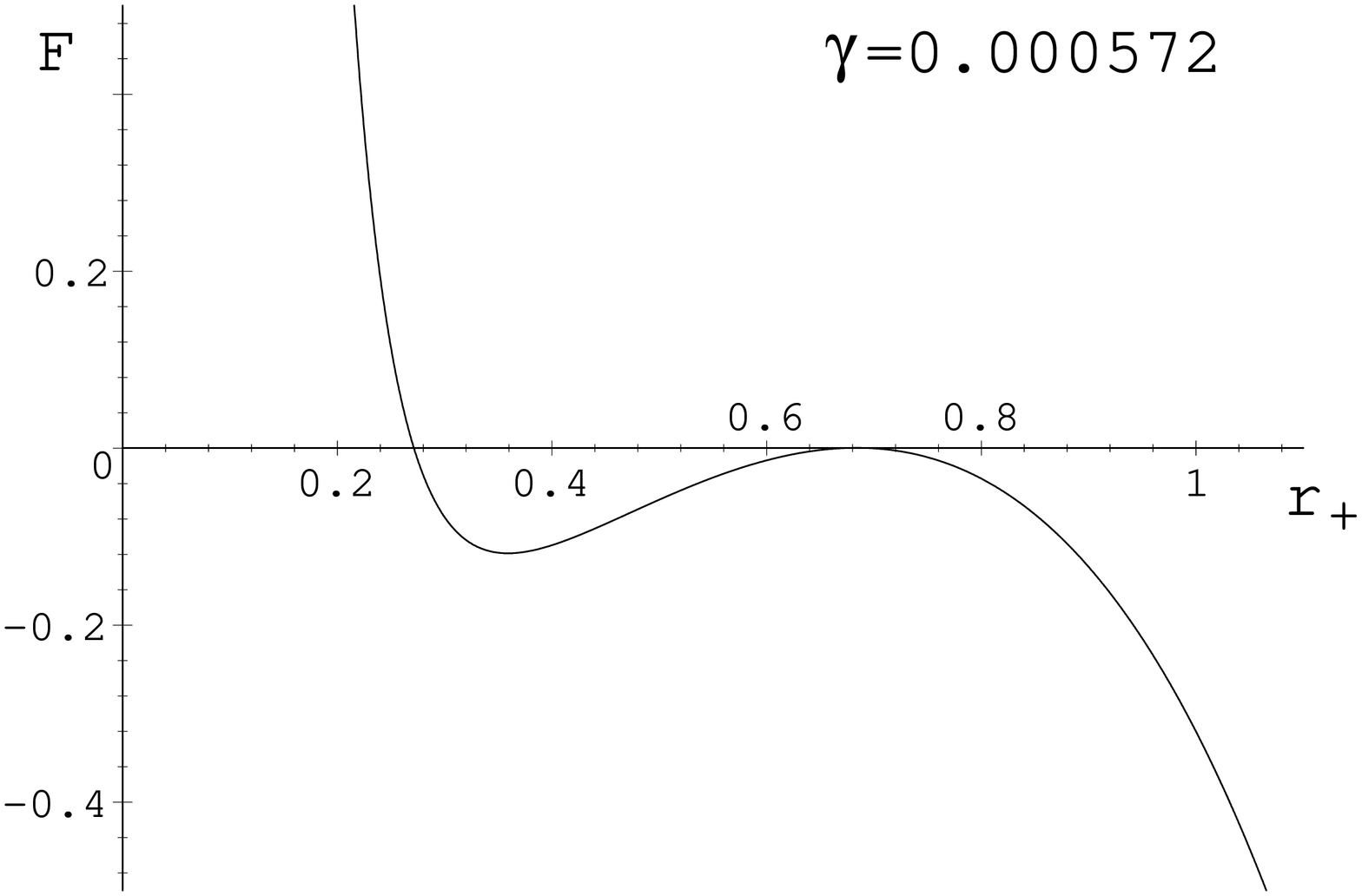,angle=-90,width=0.5\linewidth}}
\caption{{\sf Free energy $F$ in function of the horizon
position $r_+$ for spherical black holes. Two new critical values of
$\gamma$ appear here: $\gamma_1^{crit}\approx 0.000275$, where the branch
of small black holes begins to dominate over AdS space, and
$\gamma_2^{crit} \approx 0.000572$, where the AdS phase disappears.}}
\label{fig:fr}
\end{figure}

\begin{figure}[htb]
\centerline{\epsfig{file=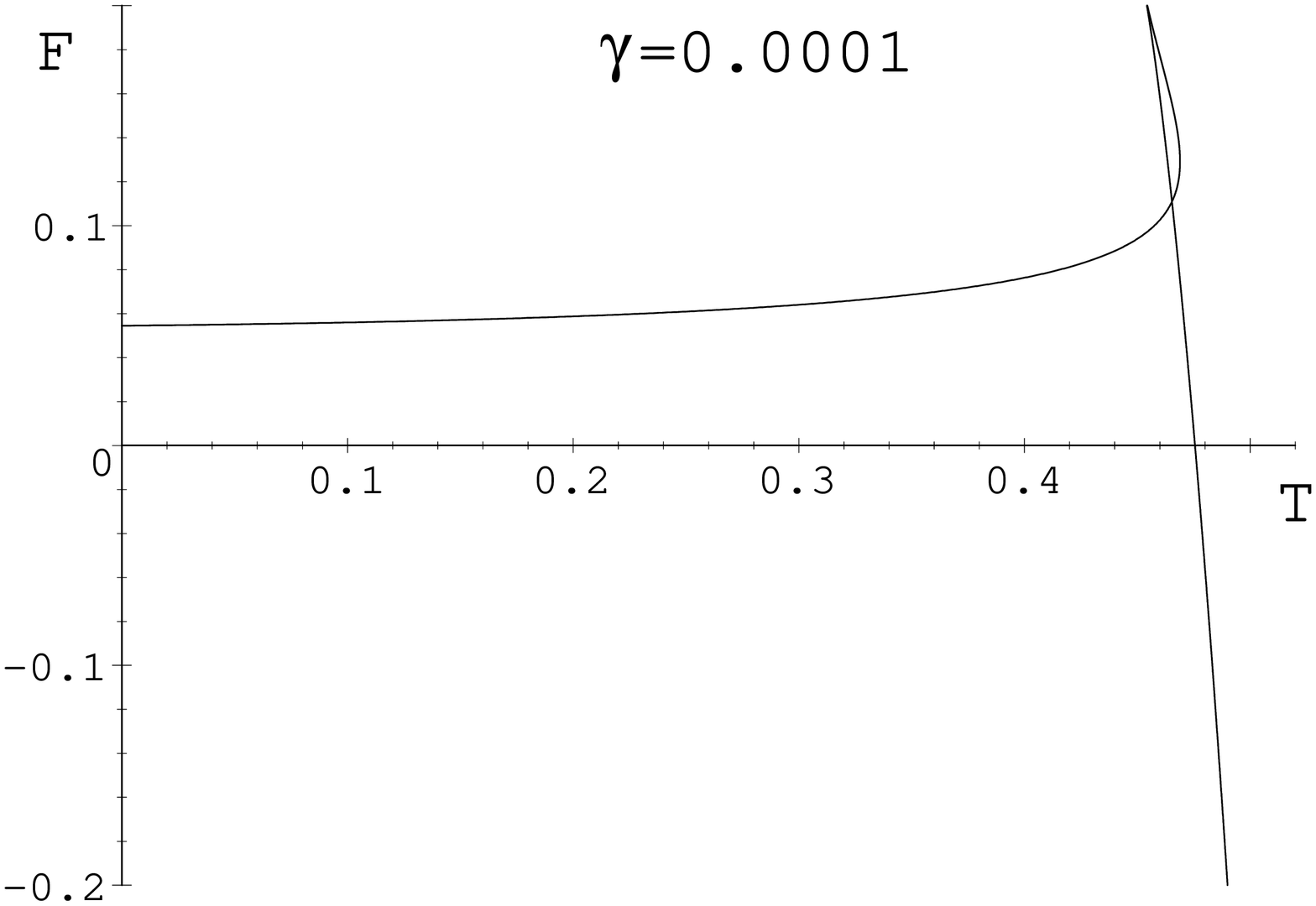,angle=-90,width=0.5\linewidth}
\epsfig{file=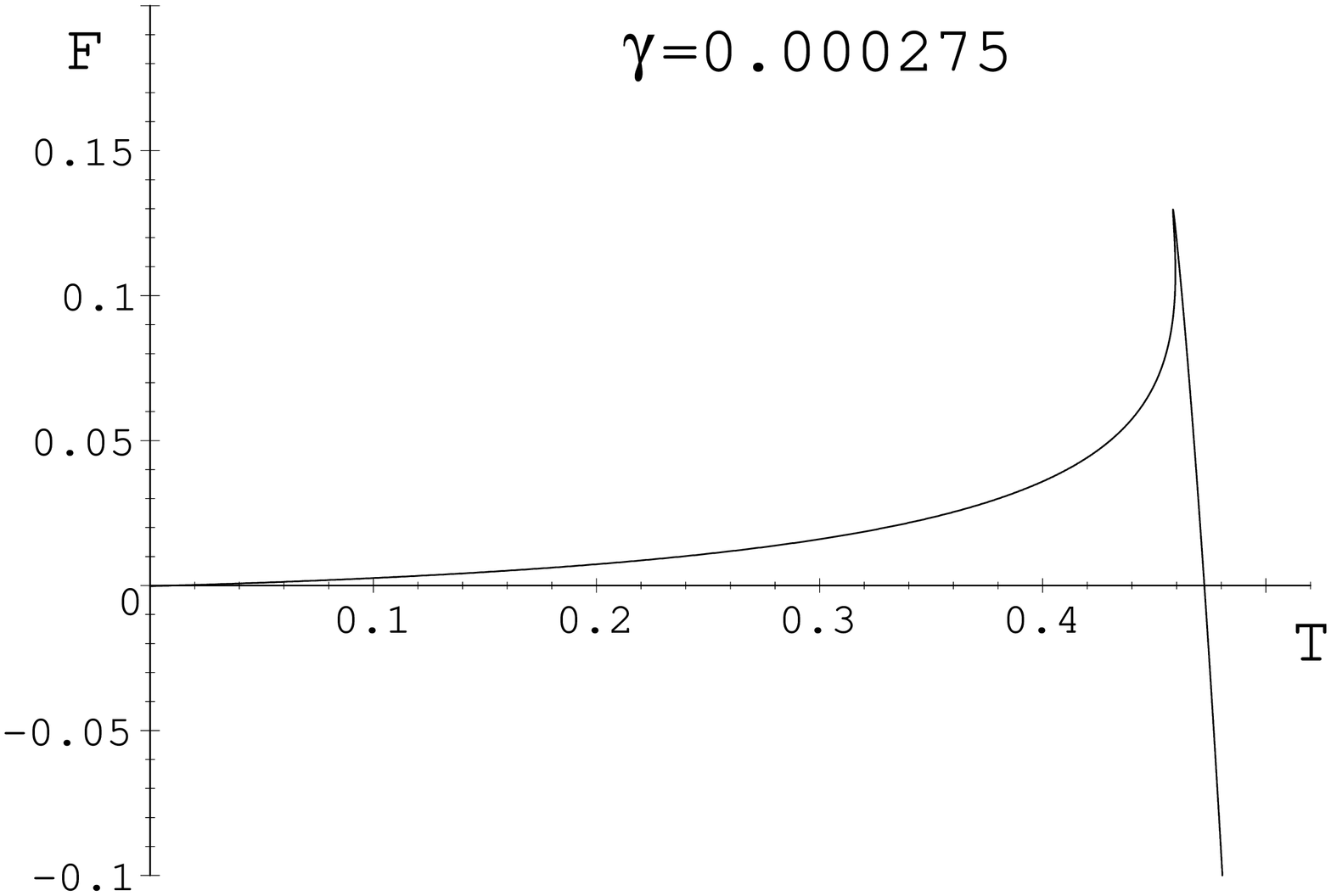,angle=-90,width=0.5\linewidth}}
\centerline{\epsfig{file=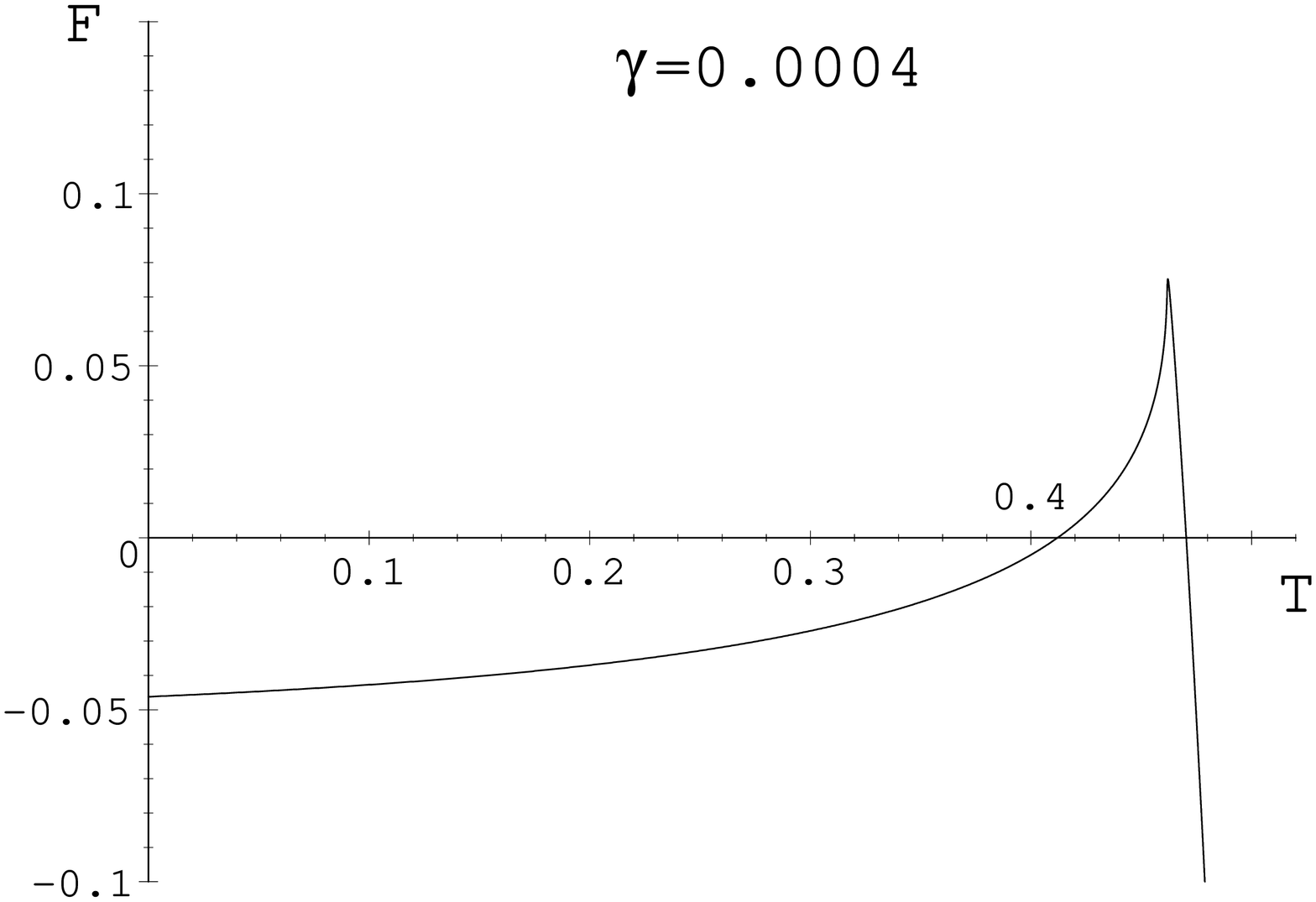,angle=-90,width=0.5\linewidth}
\epsfig{file=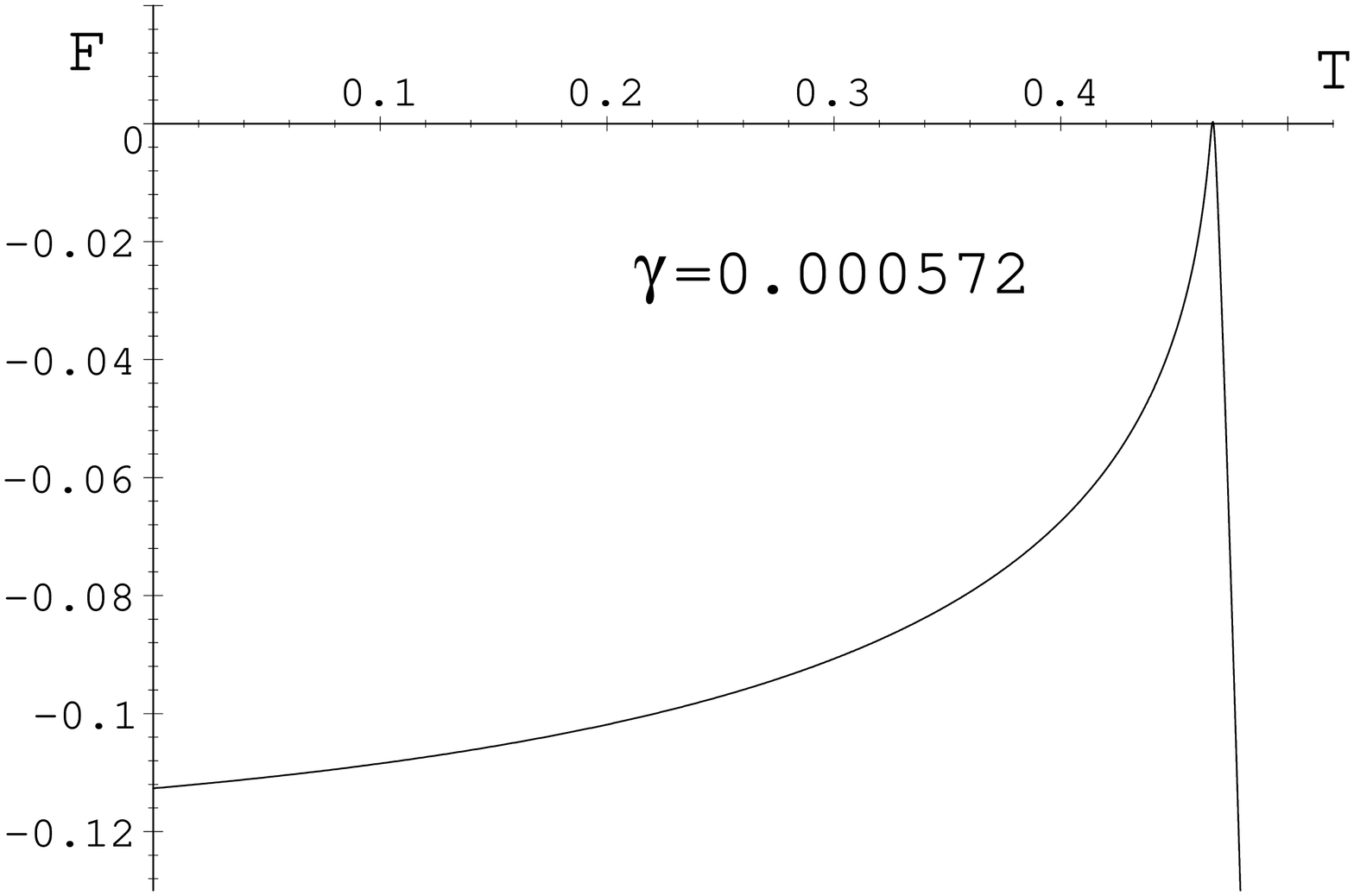,angle=-90,width=0.5\linewidth}}
\caption{{\sf Free energy $F$ in function of the temperature $T$ for
spherical black holes. For $\gamma<\gamma^{crit}_1 \approx 0.000275$
there is the standard
Hawking-Page phase transition; at $\gamma_1^{crit}$
a new low-temperature
phase of small black holes appears, and merges to the large black hole
phase at $\gamma^{crit}_2 \approx 0.000572$.}}
\label{fig:ft}
\end{figure}

\begin{figure}[htb]
\centerline{\epsfig{file=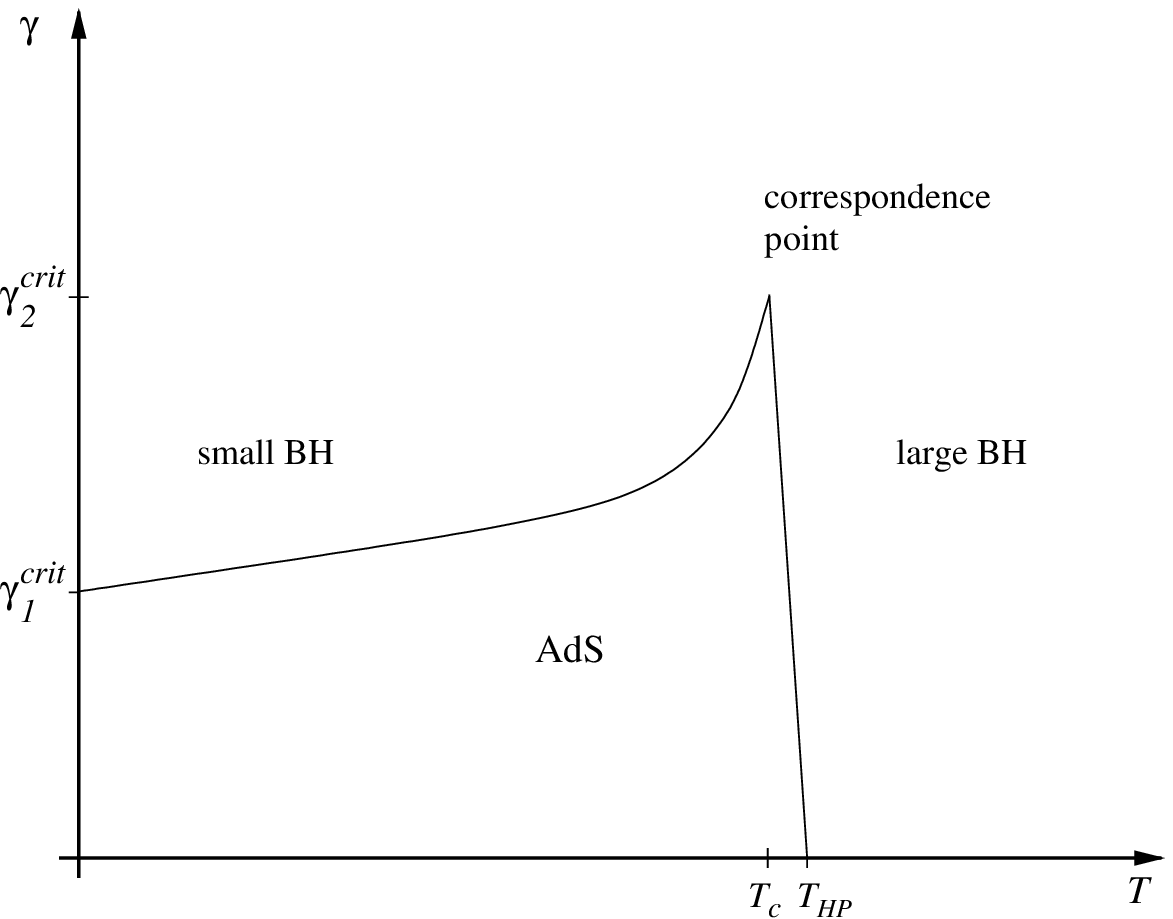,width=0.8\linewidth}}
\vskip3mm
\caption{{\sf Phase diagram for ${\cal N}=4$ SYM on $S^1 \times S^3$
for sufficiently large `t Hooft coupling $g_{YM}^2N$. 
For $\gamma<\gamma^{crit}_1$ AdS dominates over the small black holes and
we have the usual Hawking-Page transition. Note that its critical
temperature $T_{HP}$ decreases as $\gamma$ increases. 
At $\gamma_1^{crit}\approx 0.000275$, corresponding to $g_{YM}^2N
\approx 33.4$,
the small black holes begin to dominate over AdS for small $T$, and we have
three phases, a small black hole phase for small $T$, separated by an AdS phase
from the large black hole phase. At $\gamma_2^{crit}\approx 0.000572$,
corresponding to $g_{YM}^2N \approx 20.5$,
the two first order transitions degenerate to a higher order transition at
$T_c\approx 0.467/l$, and for larger $\gamma$
we are left with a unique black hole phase.}}
\label{fig:phase_d5}
\end{figure}

We now want to use the obtained thermodynamical quantities to
determine the phase diagram, which is also relevant for the
corresponding superconformal field theory via the AdS/CFT dictionary.
To begin with, we study the case $d=5$, $k=1$, which corresponds
to SYM on $S^1 \times S^3$. In order to translate our results
from the bulk to the SYM side, we will make use of the relation
\eq
\frac{\alpha'}{l^2} = (2g_{YM}^2N)^{-\frac12}.
\feq
A first important thermodynamical information is obtained observing the
behaviour of the inverse temperature $\beta$ in function of the black hole
size $r_+$. The function $\beta(r_+)$ has been plotted for different values
of the string perturbation $\gamma$ in figure~\ref{fig:br}.
We immediately see that there exists a stable black hole corresponding to
each value of the temperature, whereas in the unperturbed case there was a
minimum temperature. This is due to the emergence
of a stable branch of small black holes in the perturbed solution.
Furthermore, there is a
minimum size $r_+^{min}$ for stable black holes,
where the temperature vanishes. 
For small $\gamma$, the
small black holes are separated from the large black holes by a branch of
unstable intermediate black holes. At $\gamma_{infl}\approx 0.0004$ the
unstable branch degenerates to a point of inflection of $\beta(r_+)$,
and for larger
$\gamma$ the two stable branches merge together and no more distinction
can be made beetween the small and large black hole phases. 
All this reminds us of the behaviour of a Van~der~Waals gas in the Clapeyron
$(V,P)$ plane; and one could expect the point of inflection in
$\beta(r)$ to signal a
critical point. To go into this, however, we have to study the behaviour of
the free energy, and take into consideration that in our system a third
phase -- AdS space filled with thermal radiation -- is available.
For charged AdS black holes, an analogous behaviour has recently been
observed in \cite{cejm,cg}.\\
In figure~\ref{fig:fr} we have reported the free energy of the perturbed
solution as a function of $r_+$ for different values of $\gamma$ (in all
the figures the free energy $F$ is given in units of $V_3/(16\pi G_5)$).
To analyse these graphics, we must keep in mind that we computed the free
energy with respect to AdS space, hence positive free energy means that
$AdS$ space filled with thermal radiation is thermodynamically favoured,
while negative free energy signals the collapse of the thermal radiation
into a black hole (for a detailed analysis of stability and of the
definition of the canonical ensemble, we refer to \cite{cg}). For small
$\gamma$ we have the usual behaviour, with AdS
space thermodynamically favoured with respect to the small black hole, while
for sufficiently large $r_+$ the canonical ensemble is dominated
by the large black hole. At the critical value
$\gamma_1^{crit}\approx 0.000275$ (corresponding to $g_{YM}^2N \approx
33.4$) the free energy vanishes for 
$r_+\approx 0.36 l$, corresponding to the small black hole branch. For
larger $\gamma$, we should distinguish four cases, but considering that
there is a minimal size of the stable black holes, only three cases are
relevant.
We have first for small $r_+$ a region where the small black hole dominates,
then, as the size of the black hole grows, there is a
transition to AdS space, and
finally a second transition (the original Hawking-Page transition), to
a large black hole phase. For $\gamma^{crit}_2\approx 0.000572$
(corresponding to $g_{YM}^2N \approx 20.5$), the
region of AdS-domination disappears and for larger $\gamma$ the black hole
always dominates, if its size is greater that $r_+^{min}$. Observe that, from
the study of $\beta(r_+)$, the distinction between small and large black
holes disappeared at $\gamma_{infl}<\gamma^{crit}_2$,
whereas this distinction is
reintroduced by the existence of a third phase, AdS space, and disappears
definitively only at $\gamma_2^{crit}$.\\
We are ready now to examine the free energy as a function of $T$ and
$\gamma$, from which we shall deduce the phase structure of our system. 
The graphs of $F(T)$ for different values
of $\gamma$ are reported in figure~\ref{fig:ft}.
For $\gamma<\gamma^{crit}_1$ we have a phase dominated by AdS space filled
with thermal radiation at low temperatures, and a phase dominated by large
black holes at high temperatures, separated by the Hawking-Page
transition. For $\gamma^{crit}_1<\gamma<\gamma^{crit}_2$, a new phase
transition appears at low temperatures, separating a low-temperature phase
of small black holes from an intermediate-temperature AdS phase. At high
temperatures we have the Hawking-Page transition and again the large black
hole dominates.
Note that the critical temperature of this transition is lowered by
$\gamma$-corrections, as first shown in \cite{gaoli}.
Finally, at $\gamma^{crit}_2$ the AdS phase disappears completely, and for
higher $\gamma$ we have a unique phase dominated by black holes.
For $\gamma>\gamma^{crit}_2$ the distinction between small and large black
holes disappears; the system behaves like a liquid-vapour system. 
In the critical point $(\gamma^{crit}_2,T_c\approx 0.467/l)$ we expect a
higher order phase transition where the Hawking-Page phase transition
disappears; this point has been identified by Gao and Li with the
Horowitz-Polchinsky correspondence point in \cite{gaoli}.
The phase structure is summarized in the phase diagram of
figure~\ref{fig:phase_d5}.
We stress the fact that the critical values $\gamma_1^{crit}$ and
$\gamma_2^{crit}$ are very small (of the order of $10^{-4}$); as a
consequence there should be no further significant corrections of 
higher order in $\gamma$ in the region of the phase diagram we explored.
On the ${\cal N}=4$ SYM side, one has the correspondence point at
$g_{YM}^2N\approx 20.5$, which is considerably larger than the value of
$1.65$ obtained by Gao and Li.
The reason is that they extrapolated the (steep) Hawking-Page transition
curve to zero temperature, whereas we showed that this curve is interrupted
by the intersection with the coexistence curve of small black holes and AdS
space
at a much smaller value of $\gamma$. Hence, in the SYM theory,
the higher order phase transition is located at a value of the `t~Hooft
coupling where we have a better control on the large $N$ expansion.

\begin{figure}[htb]
\centerline{\epsfig{file=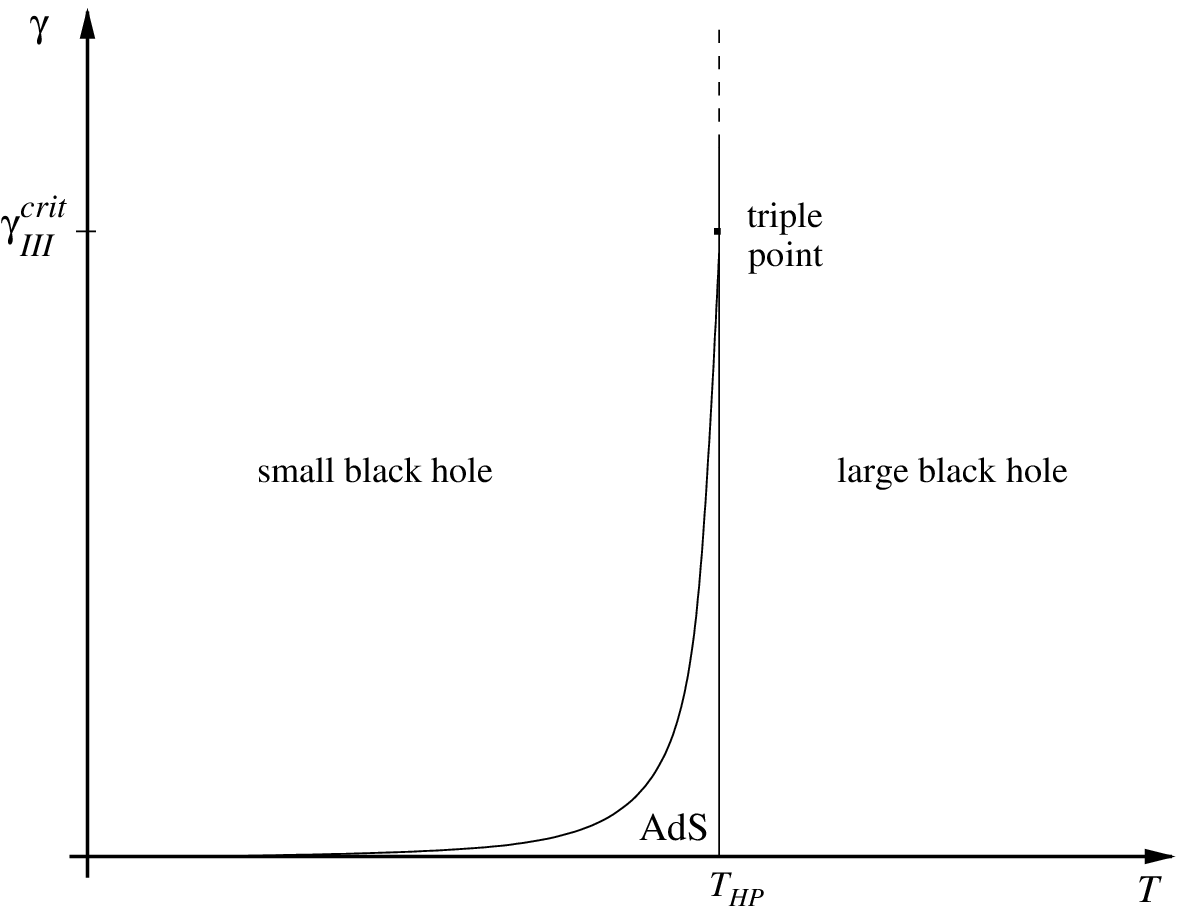,width=0.8\linewidth}}
\vskip3mm
\caption{{\sf Phase diagram for ${\cal N}=8$ SYM on $S^1 \times S^2$ with
global $SO(8)$ R-symmetry. 
For $0<\gamma<\gamma^{crit}_{III}$ there is a small black hole phase for small
$T$, separated by an AdS phase from the large black hole phase. At
$\gamma_{III}^{crit}\approx 3.07\cdot 10^{-5}$, the AdS phase disappears,
leaving a triple point ($T_c\approx 0.32/l$). For higher $\gamma$ we have a
first order transition between a small and a large black hole phase.}}
\label{fig:phase_d4}
\end{figure}

\begin{figure}[htb]
\centerline{\epsfig{file=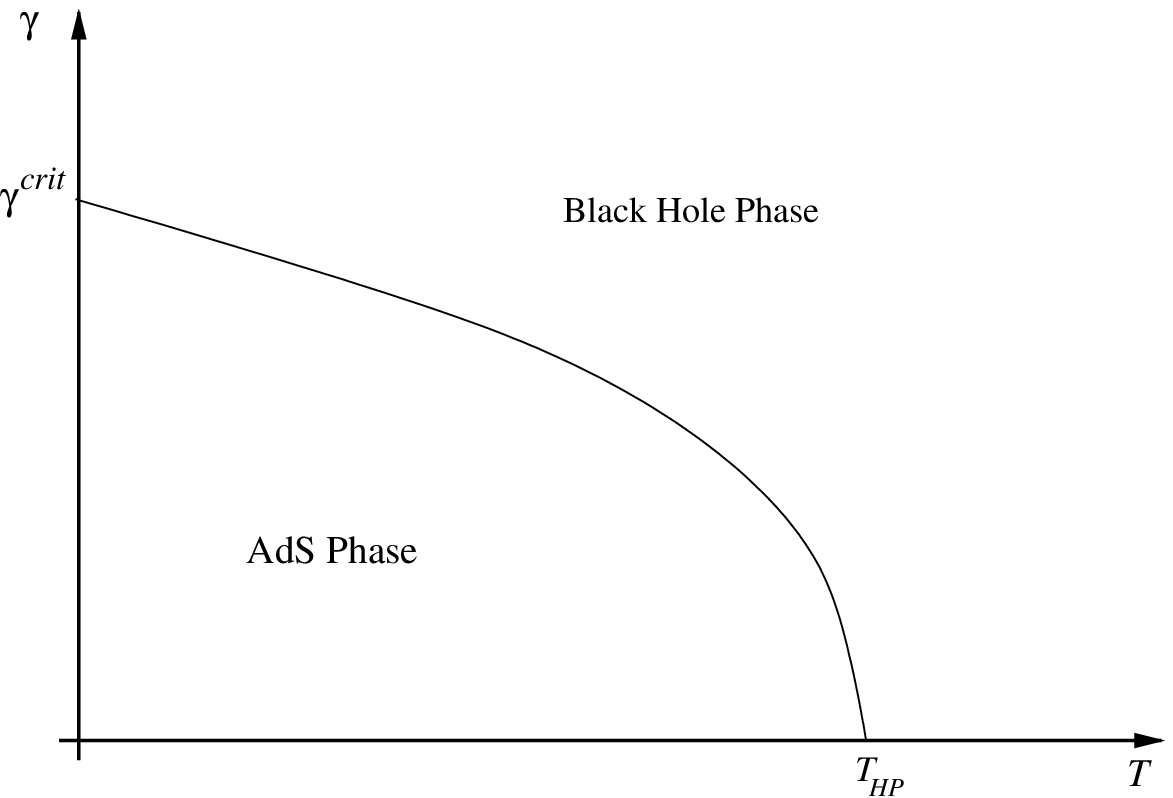,width=0.8\linewidth}}
\vskip3mm
\caption{{\sf Phase diagram for $(0,2)$ CFT on $S^1 \times S^5$.
For $\gamma<\gamma^{crit}$, AdS dominates over the black hole and
we have the usual Hawking-Page transition. However $T_{HP}$ decreases as
$\gamma$ increases.  
At $\gamma^{crit}\approx 5.48\cdot10^{-5}$,
the Hawking-Page transition disappears, leaving a unique black hole phase
for higher $\gamma$.}}
\label{fig:phase_d7}
\end{figure}

We shall now turn to the $d=4$ and $d=7$ cases. The analysis of their
thermodynamical behaviour can be carried over as in the $d=5$ case; as no
further technical problems arise, we shall not report it here and instead
discuss simply their phase structure.\\
The phase diagram for four-dimensional black holes is shown in figure~\ref{fig:phase_d4}.
For non-zero $\gamma$ a new phase of small black holes appears at low
temperature and dominates over AdS space filled with thermal
radiation. The low temperature phase is separated by a first order phase
transition from
an AdS phase. The coexistence curve small~black~hole/AdS terminates in the
origin $\gamma=0$, $T=0$. At higher temperature we have the Hawking-Page
phase transition between AdS and a phase dominated by a large black hole.
Observe that the Hawking-Page temperature remains approximately constant as
$\gamma$ grows. At $\gamma_{III}^{crit}\approx3.07\cdot10^{-5}$ the two
coexistence curves intersect for $T\approx0.32/l$ in a triple point. For
higher $\gamma$ we are left with a small/large black hole phase separation
curve, which continues more or less at the original Hawking-Page transition
temperature; this curve could end in a second critical point like in the
phase diagrams studied in \cite{cejm}, but in our perturbative approach 
we are not able to clarify this issue.
From the AdS/CFT point of view, figure~\ref{fig:phase_d4} corresponds to
the phase structure of
${\cal N}=8$ super-Yang~Mills theory on $S^1 \times S^2$
with global $SO(8)$ R-symmetry, in the limit of sufficiently large $N$.\\
The seven-dimensional black hole case is somewhat different; AdS space
always dominates over the small black hole. For this system, the phase
diagram is drawn in figure~\ref{fig:phase_d7}.   
For small $\gamma$ we have a low temperature AdS phase separated by a
Hawking-Page transition from a high temperature black hole phase. The
Hawking-Page transition temperature is lowered as $\gamma$ grows, and
vanishes for $\gamma^{crit}\approx5.48\cdot10^{-5}$; for higher $\gamma$
only the black hole phase remains. From the boundary conformal field theory
point of view, the spherical seven-dimensional AdS black hole in M-theory
compactified on a four-sphere corresponds to an exotic six-dimensional
theory with $(0,2)$ supersymmetry. The flat analogue of this theory is
particularly interesting; Witten \cite{witten2} suggested that with a
supersymmetry-breaking compactification to
four dimensions it may be possible to explore four-dimensional $SU(N)$
Yang-Mills theory in the large $N$ limit; this theory was therefore
considered one of the main approaches to confinement in QCD.

Finally, for the hyperbolic or flat cases, $k=-1$ or $k=0$, one checks that 
the corrected free energy $F(T)$ is always negative for positive $T$;
it is zero only for the background, i.~e.~for the extremal black hole
with $T=0$. This means that the (deconfining) black hole phase
dominates for all temperatures, also if one takes the leading
stringy (or eleven-dimensional Planck length) corrections into account.
So these do not cause any phase transition. However, if the transverse
manifold (i.~e.~${\mathbb H}^p$ for $k=-1$ and $\R^p$ for $k=0$)
is compactified, it supports non-contractible 1-cycles around which
strings can wind. When the length of these cycles become comparable
to the string scale $\ell_s$,
the winding modes become very light, and we expect
the geometry to be modified. Now for the hyperbolic black hole
there is a minimum horizon radius $r_0$ given by \rif{r0}. The
length of the noncontractible cycles in general depends on several
parameters, like the moduli and genus in the case of a compact Riemann surface
($p=2$). If we neglect for the moment this dependence and suppose those
lengths to be of order one, light winding modes become important for
$r_0 \sim \ell_s$, which gives a rather large value of $\alpha'$.
The behaviour is different for the toroidal black hole however, as here the
horizon radius can become arbitrarily small, it shrinks to zero size
in the extremal case $T=0$. So light winding modes can modify the
geometry already for small values of $\alpha'$. These finite size
effects have been discussed in \cite{barbon}. Let us take a short look on
what happens when the moduli of the torus are taken into account.
As a model we take the (uncorrected) black hole metric in IIB supergravity,
\eq
ds^2 = -f(r)dt^2 + f(r)^{-1}dr^2 + r^2(dx^2 + |\tau|^2dy^2 +
       2{\mathrm{Re}}(\tau)dxdy + dz^2) + l^2 d\Omega_5^2, \label{Xbh}
\feq
where
\eq
f(r) = -\frac{\eta}{r^2} + \frac{r^2}{l^2}
\feq
and $x,y,z \in [0,1]$. For the sake of simplicity, we restrict ourselves
to one complex modulus $\tau$. The fundamental cycles in $x$ or $z$
direction have length 1, while those in $y$ direction have length
$|\tau|$. Winding modes along $x,z$ then become important for
$r_+ \sim \ell_s$, and winding modes along $y$ for $r_+ \sim \ell_s/|\tau|$.
Note that these two lengths can differ considerably, if $|\tau| \gg 1$
or $|\tau| \ll 1$. Let us assume e.~g.~the latter case. Then for
$r \sim \ell_s/|\tau|$ one should go over to a T-dual language
(T-duality along $y$). The crossover temperature is
\eq
T = T_1 \equiv \frac{\ell_s}{\pi l^2 |\tau|}.
\feq
For $r \sim \ell_s$ one has to perform another T-duality along $x,z$,
where now the corresponding crossover temperature is $T = T_2
\equiv T_1|\tau| \ll T_1$. In each of the three regimes,
$0 \le T \le T_2$, $T_2 \le T \le T_1$, and $T > T_1$, one now
has to examine the thermodynamic stability of the corresponding
manifold, as has been done in \cite{barbon} for $\tau = i$, i.~e.~for
$T_1=T_2$.
In this case, the authors of \cite{barbon} argued that there is a
phase transition at a temperature
\eq
T_{trans} = (2g_{YM}^2N)^{-\frac{1}{4}}T_1 = \frac{\ell_s}{l}T_1,
\feq
below which the dual $\tilde{X}_{bh}$ of the black hole manifold
$X_{bh}$ \rif{Xbh}, which represents a system of D0-branes smeared over the
torus $T^3$ (cf.~\cite{barbon} for the T-dual metric), goes over to
a system of D0-branes localized on the $T^3$. This phase transition
is the toroidal analogue of the Hawking-Page transition which occurs
for spherical black holes. If the modulus $|\tau|$ is very different from
$i$ (so that e.~g.~$T_1 \gg T_2$), we expect a more complicated
phase diagram, showing perhaps more than one phase transition.
(A full thermodynamical discussion involving the determination
of the phase diagram would go beyond the scope of this paper, so we
leave this issue for future investigations). In particular, we expect
a rather complicated phase diagram in the moduli space of the torus.

\section{Summary and Discussion} \label{disc}

In the present paper, we computed stringy or M-theory corrections
to $d$-dimensional anti-de~Sitter black holes for $d=4,5,7$.
Thereby we allowed for spherical, flat or hyperbolic event horizons.
In order to obtain the corrections to the black hole thermodynamics,
we first used a simplified perturbative approach, which consists in
plugging the unperturbed metric into the term quartic in the Weyl
tensor appearing in the modified supergravity action.
Then also the corrected black hole metrics have been determined explicitely
in the various cases. Quite surprisingly, the resulting thermodynamical
quantities, like free energy or entropy, coincide with those obtained
by the above mentioned simplified approach. This coincidence holds
for all values of $d$ and for all event horizon topologies under
consideration. Especially in the hyperbolic case, this agreement
is remarkable, as the calculation of the action for hyperbolic
black holes involves a nontrivial background subtraction.
For black holes with flat or hyperbolic event horizons, we found that
the first stringy or M-theory corrections do not give rise to any phase
transition; the system is in the black hole phase for any temperature.
If the spatial black hole geometry supports noncontractible 1-cycles,
i.~e.~if one compactifies the $\R^p$ or the ${\mathbb H}^p$, we have
winding modes which become very light if the length of these 1-cycles is
of order the string scale. Then we expect the geometry to be modified.
For hyperbolic black holes, which have a minimum event horizon radius
of the order the AdS scale $l$, these winding modes in general become
light only for a large value of $\alpha'$. (Taking into account
the moduli of the compact space could change this behaviour, but we
did not consider them here). For toroidal black holes however, there
is no minimum event horizon radius, so already for
small values of $\alpha'$ light winding modes can become important.
As these finite size effects have already been discussed in \cite{barbon}
for symmetric tori, we limited ourselves to a short discussion of the
nonsymmetric case, indicating the appearance of a rather complicated
phase diagram.\\
As for spherical black holes, which are relevant to study SYM on
$S^1 \times S^p$, we showed that the stringy or M-theory corrections
give rise to the emergence of a stable branch of small black holes.
By considering the free energy, the corresponding SYM phase diagram
has been determined for $d=5$ in the limit of sufficiently large
`t Hooft coupling, where the restriction to the ${\cal O}(\alpha'^3)$
corrections is valid. The phase diagram shows that the
Hawking-Page transition temperature decreases with increasing string scale,
and finally disappears at a value of $g_{YM}^2N \approx 20.5$,
which is considerably larger than the estimate given by Gao and
Li \cite{gaoli}. This critical point in the phase diagram has been
identified by the authors of \cite{gaoli} with the correspondence
point of Horowitz and Polchinski.
We found that there is a second critical
value of the `t Hooft coupling in the $S^1 \times S^3$ SYM
phase diagram, namely $g_{YM}^2N \approx 30.4$. Above this value, we
have the confining phase of thermal AdS space for low temperatures and the
deconfining black hole phase for sufficiently high temperatures.
For $20.5 < g_{YM}^2N < 30.4$, one has a phase of small black holes for
low temperatures, an AdS phase in an intermediate temperature regime,
and finally a phase of large black holes for high temperatures.
For $g_{YM}^2N < 20.5$, the system is always in the black hole phase;
besides, the small and large black holes have become indistinguishable.
The $d=4$ black hole, corresponding to ${\cal N}=8$ SYM on $S^1 \times S^2$,
has a somewhat different phase structure. The phase of small black holes
now already appears for any nonzero value of $\gamma$. Furthermore, the
critical point is a triple point. Above it, no AdS phase appears, but
there is still a small/large black hole phase transition.
Finally, for $d=7$ (corresponding to an exotic $(0,2)$ CFT on 
$S^1\times S^5$), there is no small black hole phase, and the AdS phase is
located in the region of small temperatures and small values of $\gamma$.

\section*{Acknowledgements}

The authors would like to thank G.~Cognola, F.~Ferrari, M.~Porrati and
L.~Vanzo for helpful discussions.\\
The part of this work due to D.~K.~has been supported
by a research grant within the
scope of the {\em Common Special Academic Program III} of the
Federal Republic of Germany and its Federal States, mediated 
by the DAAD.

\end{document}